\documentclass[usenatbib,onecolumn]{mn2e}
\usepackage{amsmath}
\usepackage{graphicx}
\usepackage{enumerate}
\usepackage{subfigure}
\usepackage[T1]{fontenc} 
\usepackage{aecompl}
\usepackage{verbatim}

\newcommand{\VEC}[1]{\mathbf{#1}}

\title{Analytic model for the matter power spectrum, its covariance 
matrix, and baryonic effects} 

\author[I. Mohammed et al.] {Irshad
  Mohammed,\thanks{irshad@physik.uzh.ch}$^1$ and Uro\v s Seljak$^2$ \\ \\
  $^1${Physik-Institut, University of Zurich, Winterthurerstrasse 190, 8057 Zurich, Switzerland} \\ 
  $^2${Department of Physics, Department of Astronomy and Lawrence Berkeley National Laboratory,}\\
	{  University of California, Berkeley, California 94720, USA}
}

\begin{document}
\maketitle
\begin{abstract}

We develop a model for the matter power spectrum as the sum of Zeldovich approximation and even powers of $k$, i.e., $A_0 - A_2k^2 + A_4k^4 - ...$, compensated at low $k$. With terms up to $k^4$ the model can predict the true power spectrum to a few percent accuracy up to $k\sim 0.7 h \rm{Mpc}^{-1}$, over a wide range of redshifts and models. 
The $A_n$ coefficients  contain information about cosmology, in particular amplitude of fluctuations.  
We write a simple form of the covariance matrix as a sum of Gaussian part and $A_0$ variance, which reproduces the simulations remarkably well. In contrast, we show that one needs an N-body simulation volume of more than 1000 $({\rm Gpc}/h)^3$ to converge to 
1\% accuracy on covariance matrix. We investigate the super-sample variance effect and show it can be modeled as an additional parameter that can be determined from the data. This allows a determination of $\sigma_8$ amplitude to about 0.2\% for a survey volume of 1$({\rm Gpc}/h)^3$, compared to 0.4\% otherwise.
We explore the sensitivity of these coefficients to baryonic effects using hydrodynamic simulations of \cite{2011MNRAS.415.3649V}. We find that because of baryons redistributing matter inside halos all the coefficients $A_{2n}$ for $n>0$ are strongly affected by baryonic effects, while $A_0$ remains almost unchanged, a consequence of halo mass conservation. Our results suggest that observations such as weak lensing power spectrum can be effectively marginalized over the baryonic effects, while still preserving the bulk of the cosmological information contained in $A_0$ and Zeldovich terms.

\end{abstract}


\section{Introduction}\label{sec:intro}

The clustering of dark matter as a function of scale and redshift
contains useful information about many cosmological parameters. 
For example, clustering as a function of redshift is very sensitive to the 
dark energy density and its equation of state.
Clustering as a function of scale can reveal information about the primordial slope of the power spectrum and
matter density, as well as about the presence of massive neutrinos. 
The best way to measure the dark matter clustering is via weak lensing \citep{2001PhR...340..291B,2003ARA&A..41..645R}. 
In weak lensing light from distant galaxies, called sources, is being deflected by mass distribution along the 
line of sight, such as that the images are distorted. The primary distortion is shear, which changes ellipticity 
of the light of the source galaxy. By correlating these ellipticities between the source galaxies one can deduce 
the clustering strength of the matter along the line of sight. 
Over the past decade 
this recognition put weak lensing surveys at the forefront of cosmological probes, with several 
ground based and space based experiments proposed \citep{2006ApJ...647..116H,2007MNRAS.376...13M, 2008A&A...479....9F, 2010A&A...516A..63S}. 
The primary statistic is the convergence power spectrum $C_l^{\kappa \kappa}$, which can be expressed as a weighted 
projection over the matter power spectrum $P(k)$ along the line of sight from the observer to the source. 
Future surveys will contain sources at many different redshifts, and by combining this information one 
can minimize the line of sight projection and measure a quantity close to the 3-dimensional 
power spectrum, a procedure called weak lensing tomography. In this paper we will focus on 
the 3-dimensional power spectrum of matter $P(k)$. 

The procedure to extract information from the weak lensing measurements 
is in principle straight-forward and, while experimentally challenging, its theoretical 
underpinnings have been known for a long time. 
What are the remaining theoretical challenges in this program? The predictions of the 
dark matter only (DMO)  clustering on small scales, where nonlinear effects are important, was one of the 
uncertainties. For example, 
the widely used HALOFIT \citep{2003MNRAS.341.1311S} is only accurate to 10\%, although
the revised version \citep{2012ApJ...761..152T} is argued to be $5\%$ accurate for $k < 1.0\ h \rm{Mpc}^{-1}$. 
Recent progress in N-body simulations suggests this problem will soon be solved.  
For example, the \textit{The Coyote Universe} DMO power spectrum 
emulator \citep{2010ApJ...715..104H,2009ApJ...705..156H,2010ApJ...713.1322L} is accurate to nearly 1$\%$ up to $k\sim 1\ h \rm{Mpc}^{-1}$ for the 38 cosmologies that have been simulated. The emulator provides an output power spectrum for 
any cosmological model, interpolated from the grid of 38 simulated models, 
with an error that can be as high as 5$\%$ for some cosmological models. 
It seems likely that the precision will reach the required level in the near future as finer grids of simulations 
are developed, but it is also clear that by using better ways to interpolate between the models could 
improve the accuracy. 

The second problem are the baryonic effects. Baryons differ from the dark matter in several aspects. 
First difference is that hot baryonic gas has pressure, which prevents clustering on small scales. These effects are particularly important 
inside the dark matter halos, where gas temperature is high and pressure effects large. In addition, 
baryons cool and condense into stars, possibly bringing dark matter along in the process. However, baryons also
form stars, which in turn lead to supernovae that can produce energy outflows. Even more dramatic effects
can arise from the active galactic nuclei (AGN), which can also produce massive energy outflows.
Recent studies with hydrodynamical simulations \citep{2011MNRAS.415.3649V} have argued that 
these AGN feedback models are required to match the observations of X-ray groups and clusters, specially the 
temperature-luminosity relation in X-rays. The outflowing baryons can also redistribute the dark matter. 
Recent work \citep{2011MNRAS.417.2020S,2013MNRAS.434..148S}
shows that the baryonic correction in the matter power spectrum can be
important above $k \sim 0.3 \ h \rm{Mpc}^{-1}$ and if one does not take account for it, 
it will bias the cosmological constraints such as dark energy equation of state \citep{2011MNRAS.417.2020S}. 

Third theoretical problem that remains unsolved
is the issue of reliable covariance matrix for the observed power spectrum and optimal weighting of the data. 
The full covariance matrix consist of two parts: Gaussian and non-Gaussian. Both scale inversely with the volume of the 
survey. Gaussian contribution is very large at large scales (low wavemodes $k$)
due to sample variance, i.e. finite number of long wavelength 
Fourier modes sampled in a finite volume. At higher $k$ the sampling variance becomes small and Fisher matrix calculations based 
on Gaussian variance have predicted that most of the cosmological information in weak lensing comes from small scales. 
However, the non-Gaussian part becomes important on smaller scales and makes these predictions unreliable. 
There are two essential contributions to the covariance matrix: 
one arises from the Poisson fluctuations in the number of halos relative to the average, and the 
second arises from the fluctuations on the scale of the survey, which induce curvature type effects
that couple to all modes inside the survey \citep{2011JCAP...10..031B,2013PhRvD..87l3504T}. For weak lensing applications these contributions become
significant for $\sim \ell > 500$ \citep{2012PhRvD..86h3504Y}. 
So far the predictions have relied either on the halo model \citep{2013PhRvD..87l3504T} 
or on the simulations \citep{2009ApJ...701..945S,2011ApJ...734...76S,2014PhRvD..89h3519L}. It has 
been argued that large numbers of simulations are needed to converge for a single model \citep{2011ApJ...734...76S,2014arXiv1406.2713B}. 
Without a reliable covariance matrix one cannot optimally combine the different power spectrum estimates, 
nor can one reliably estimate the errors, as emphasized in recent work 
\citep{2014MNRAS.442.2728T,2014MNRAS.439.2531P}. 

In this paper we propose a different approach to the dark matter power spectrum description that addresses 
all of the challenges above. We propose a novel form of the 
halo model for the dark matter power spectrum 
\citep{2000MNRAS.318..203S,2000MNRAS.318.1144P,2000ApJ...543..503M,2002PhR...372....1C}, in which we split the power spectrum into the quasi-linear 2-halo 
term, which we take to be the Zeldovich 
approximation, and the 1-halo term. Rather than relying on the analytic forms for the 1-halo term as in 
the original halo model \citep{2000MNRAS.318..203S,2000ApJ...543..503M} we simply expand 
it into the series of even powers of $k$ and fit each coefficient to the simulations. 
By doing so we obtain an accurate description of the dark matter power spectrum up to $k \sim 0.7\ h \rm{Mpc}^{-1}$. 
We then investigate the 
baryonic effects on these coefficients and address the question how to marginalize against these effects. Finally, 
the resulting solution we propose also simplifies the question of the covariance matrix calculations. 

The outline of the paper is as follows: In section \ref{sec:theory}, we review some important theoretical background, particularly the halo model (section \ref{sec:halomodel}) and Zeldovich approximation (section \ref{sec:zeldovich}). We postulate the necessary modifications in the 1-halo term in section \ref{sec:1haloexpansion} and calibrate the fitting functions on simulations in section \ref{sec:calibration} and showing the comparison with the true matter power spectrum. In section \ref{sec:covariancematrix} we discuss the covariance matrix and cosmological information content of our model. We are also discussing super-sample variance in section \ref{sec:ssv}. In section \ref{sec:baryons}, we describe the same method with baryons and the limits to which one can calculate the non-linear matter power spectrum and its full covariance matrix using this methodology. Finally in section \ref{sec:discussion}, we summarise and discuss the possibility of the future work.

\section{Theoretical model for dark matter power spectrum}
\label{sec:theory}

\subsection{The halo model}\label{sec:halomodel}

There are several approaches to account for clustering of dark matter and its evolution in the Universe. One of the 
more successful frameworks is the halo model \citep{1977ApJ...217..331M,2000MNRAS.318..203S,2000ApJ...543..503M,2000MNRAS.318.1144P,2002PhR...372....1C}. 
We will first review the halo model as implemented in previous work before presenting a new version of the halo model
that is more accurate. 
In the halo model approach, all the matter in the Universe is assumed to be in isolated halos with mass
defined by a threshold density as:
\begin{equation}
	M_{\bigtriangleup} = \dfrac{4}{3} \pi R^3_{\bigtriangleup} \bigtriangleup \bar{\rho}_m,
\end{equation}
\\
where, $M$ is the mass of the halo inside the radius $R_{\bigtriangleup}$ and the density of the halo is $\bigtriangleup$ times $\bar{\rho}_m$, which is the mean matter density of the universe. We use $\bigtriangleup=200$ throughout this paper unless stated otherwise. The power spectrum can be split into two parts:

\begin{equation}
	P(k) = P_{1h}(k) + P_{2h}(k),
\end{equation}
\\
where, the two terms in right are the 1-halo and 2-halo term respectively. The 2-halo term gives the correlation between different halos, also referred as halo-halo term, whereas the 1-halo term describe the correlation between dark-matter particles within the halo, also referred to as Poisson term, and dominates at smaller scales. These two terms are given by:

\begin{equation}
	P_{1h}(k) = \int d\nu f(\nu) \dfrac{M}{\bar{\rho}} |u(k|M)|^2, 
	\label{eqn:pk1h}
\end{equation}
\begin{equation}
	P_{2h}(k) = \left[ \int d\nu f(\nu) b(\nu) u(k|M) \right]^2 P_{\rm L}(k),
\end{equation}
\\
where, $P_{\rm L}(k)$ is the linear power spectrum. 
Throughout this paper we use publicly available code CAMB \citep{Lewis:1999bs} to compute linear matter power spectrum, unless stated otherwise. 
We also used publicly available code CHOMP\footnote{ http://code.google.com/p/chomp/} to compute some functions like the halo 
mass function and density profiles. 
The Fourier transform of the density profile of the halos $u(k|M)$ is normalized such that $u(k=0|M)=1$,
\begin{equation}
	u(k|M) = \dfrac{4 \pi}{M} \int_0^{R_{\rm vir}}dr\  r^2\  \rho(r|M)\ \dfrac{\sin(kr)}{kr}.
\end{equation}
\\
One can can see that upon expanding $\sin(kr)/kr$ only even powers of $k$ will be present, as further developed below. 
The functions $f(\nu)$ and $b(\nu)$ are the mass function and halo bias respectively. Both variables $\nu$ and $M$ account for the scale and related as:

\begin{equation}
	\nu(M,z) = \left(\dfrac{\delta_c}{\sigma(M,z)}\right)^2,
\label{nu}
\end{equation}
\\
where $\delta_c \sim 1.68$,
\begin{equation}
	\sigma^2(M,z) = \sigma^2(M)  D^2_+(z),
\end{equation}
\begin{equation}
	\sigma^2(M) =\dfrac{1}{2\pi^2} \int dk\ k^2\ P_{\rm L}(k)\ |\bar{W}(kR)|^2,
\end{equation}
\\
with, $D_+(z)$ as the growth factor and $\bar{W}(x)$ as the Fourier transform of the top-hat function:
\begin{equation}
	\bar{W}(x) = 3 \dfrac{\sin(x) - x\cos(x)}{x^3}.
\end{equation}
\\

\subsection{The new 2-halo term: Zeldovich approximation}\label{sec:zeldovich}

The halo model is not sufficiently accurate for the one percent precision required from the 
future surveys. The 2-halo term needs to be modified because in the halo model 
it is essentially given by the linear theory, and 
the nonlinear effects such as the smearing of baryonic acoustic
oscillations (BAO) are ignored. A useful improvement is the Zeldovich approximation (ZA) \citep{1970A&A.....5...84Z}. 
In it we assume the particles stream along the initial trajectory, without being perturbed by the 
nonlinear effects. Even though the Zeldovich approximation is in a sense linear, its effects on the 
density extend beyond linear effects, and ZA can even lead to caustics where the density is infinite. 
While ZA produces too little power
to be a good approximation for the fully nonlinear power spectrum, it smears the BAO
in the amount that matches the simulations quite well \citep{1993cvf..conf..585T,2008PhRvD..77f3530M}. 
As such it is a useful extension of the linear power spectrum. 
Here we will consider 
ZA approximation for large scales, coupled to the 1-halo term for the small scales. 

The Zeldovich power spectrum is given by (see e.g. \cite{1995MNRAS.273..475S})
\begin{align}
(2\pi)^3&\delta^D (k)+P(k)=\int d^3 q ~e^{-i\VEC{q}\cdot\VEC{k}}\nonumber\\
&\times\exp\left[-\frac{1}{2}k_ik_jA_{ij}(\VEC{q}))\right],
\label{eq:PSwithAW}
\end{align}
where 
\begin{equation}
A_{ij}(\VEC{q})=X(q)\delta^K_{ij}+Y(q)\hat{q}_i\hat{q}_j,
\end{equation}
and
\begin{align}
  X(q) =& \int_0^\infty \frac{dk}{2\pi^2} P_L(k)
  \left[\frac{2}{3} - 2 \frac{j_1(kq)}{kq}\right] , \\
  Y(q) =& \int_0^\infty \frac{dk}{2\pi^2} P_L(k)
  \left[-2 j_0(kq) + 6 \frac{j_1(kq)}{kq}\right] .
  \label{eq:XYex}
\end{align}
Here $P_L(k)$ is the linear power spectrum and $j_n$ is the spherical Bessel function of order $n$. 

\subsection{The 1-halo term expansion}\label{sec:1haloexpansion}
\label{sec:taylorexpansion}
In this section we first motivate the 1-halo term expansion into even powers of $k$. In the next section we 
analyze their dependence on the cosmological parameters and compare against the predictions of the halo model. 

We begin by writing the ansatz for the 1-halo term, 
\begin{equation}
	P_{1h}(k) = (A_0 - A_2k^2 + A_4k^4- ...)F(k).
	\label{eqn:pk1h_powerlaw}
\end{equation}
To motivate the ansatz and 
calculate the coefficients $A_n$, we start with the Fourier transform of the normalised density profile, assuming 
for now $F(k)=1$:
\begin{equation}
	u(k|M) = \dfrac{4 \pi}{M} \int_0^{R_{\rm vir}}dr\  r^2\  \rho(r|M)\ \dfrac{\sin(kr)}{kr}.
\end{equation}
The halo profile is spherically averaged and
assumed to depend only on the mass of the halo.
We can model the halo density profile in
the NFW form \citep{1997ApJ...490..493N}
\begin{equation}
\rho(r|M)={\rho_s \over (r/R_s)(1+r/R_s)^2}.
\label{rho}
\end{equation}
This model assumes that the profile shape is
universal in units of scale radius $R_s$, while its characteristic density
$\rho_s$ at $R_s$ or concentration $c=R_{\rm vir}/R_s$ may depend on the halo mass $M$.

The function $\sin(kr)/kr$ can be expand as Taylor series with even powers of $kr$ as
\begin{equation}
	u(k|M) = \dfrac{4 \pi}{M} \int_0^{R_{\rm vir}}dr\  r^2\  \rho(r|M)\ \left[ 
			1 - \dfrac{k^2r^2}{3!} + \dfrac{k^4r^4}{5!} - ... \right].
\end{equation}
We can simplify this equation using function $\Im_n$ as:
\begin{equation}
	u(k|M) = \Im_0 k^0 - \Im_1k^2 + \Im_2k^4-... \equiv (-1)^{n}\sum_{n=0}^{\infty} \Im_n k^{2n},
\end{equation}
and
\begin{equation}
	|u(k|M)|^2 = (-1)^{m+n} \sum_{(m,n)} \Im_n k^{2n}  \Im_m k^{2m} = 
	(-1)^{m+n} \sum_{(m,n)}\Im_m \Im_n k^{2(m+n)},
	\label{eqn:uk2}
\end{equation}
where,
\begin{equation}
	\Im_n = \dfrac{4\pi}{(2n+1)!M} \int_0^{R_{\rm vir}} dr\ r^{2(1+n)}\ \rho(r|M).
\end{equation}
Note that the functions $\Im_n$ are the integrals over the density profiles and some power of $r$ 
from 0 to $R_{\rm vir}$ and that $\Im_0=1$. However, there is nothing obviously special about truncating the 
integral there, and it can be 
changed to truncate the density profile at a different $R_{max}$ than $R_{\rm vir}$, for example $2R_{\rm vir}$. 
This suggests that the halo model has some flexibility in its implementation and is not fully predictive. 
For this reason we will just use it as a motivation and will not be doing the actual integrals over the 
halo profiles. 

Next we insert equation \ref{eqn:uk2} into 1-halo term expression of equation \ref{eqn:pk1h} and group 
the terms in even powers of $k$,
\begin{equation}
	P_{1h}(k) = \int d\nu f(\nu) \dfrac{M}{\bar{\rho}} \sum_{(m,n)}\Im_m \Im_n k^{2(m+n)},
\end{equation}
\begin{equation}
	P_{1h}(k) = \int d\nu f(\nu) \dfrac{M}{\bar{\rho}} \left[  
				\Im_0\Im_0 k^0 - 2\Im_0\Im_1 k^2 + (\Im_1\Im_1 + 2\Im_0\Im_2)k^4  - ...  
				\right]
	\label{eqn:pk1h_expand}
\end{equation}
Comparing equation \ref{eqn:pk1h_powerlaw} and \ref{eqn:pk1h_expand}, we obtain the coefficients and their variances as:

\begin{figure*}
\centering
\subfigure{\includegraphics[width=0.48\textwidth]{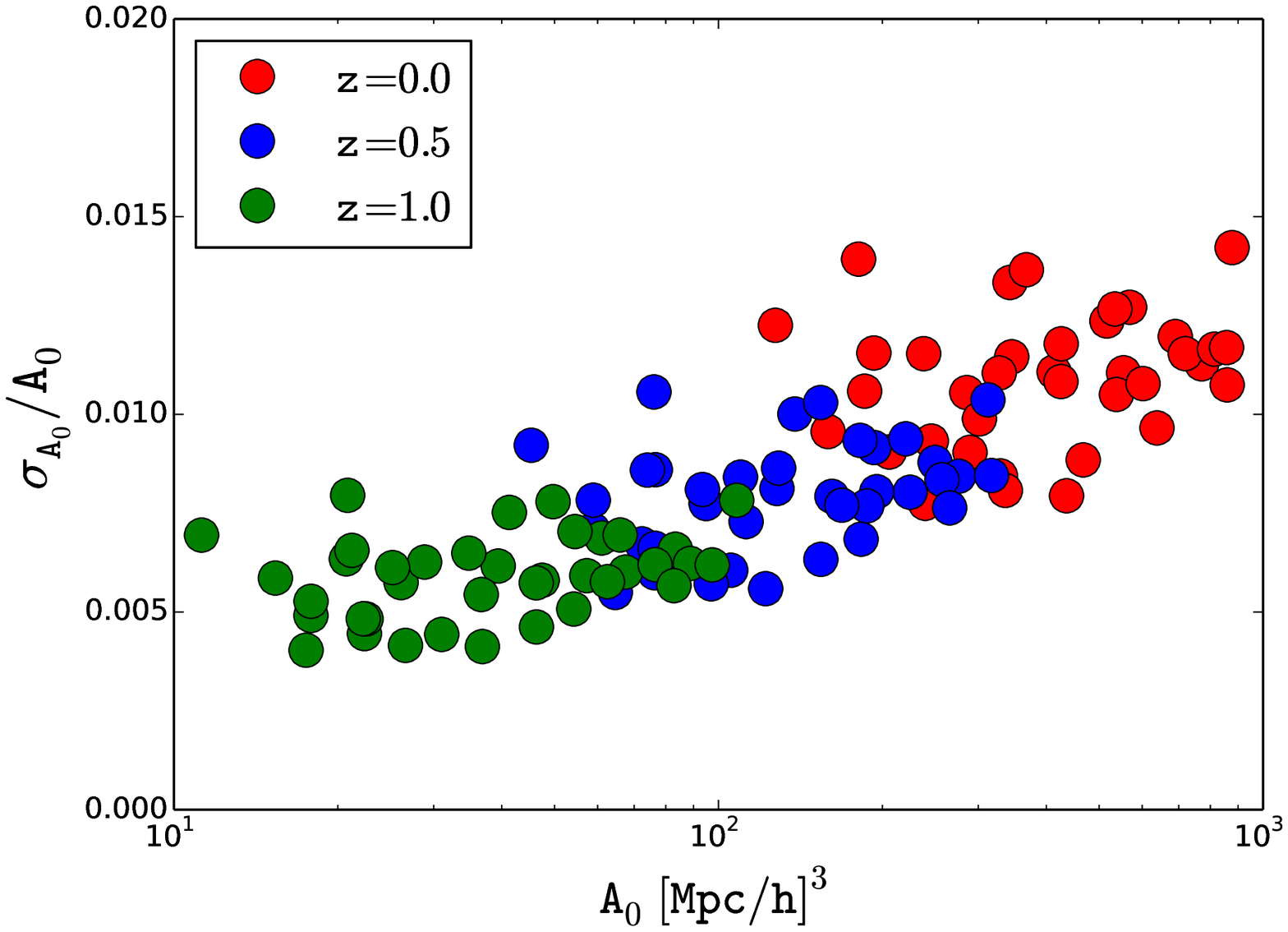}}
\subfigure{\includegraphics[width=0.48\textwidth]{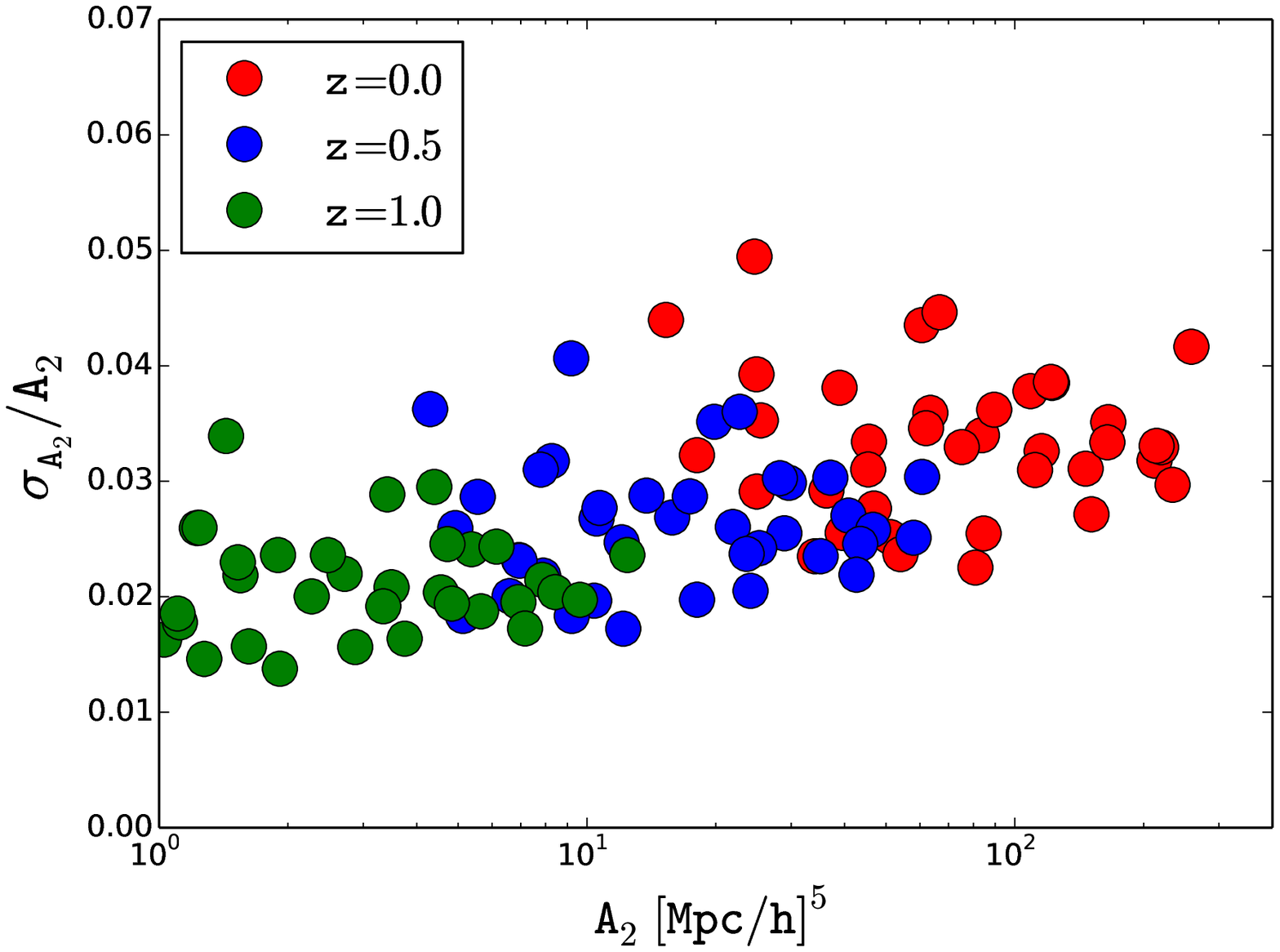}}
\subfigure{\includegraphics[width=0.48\textwidth]{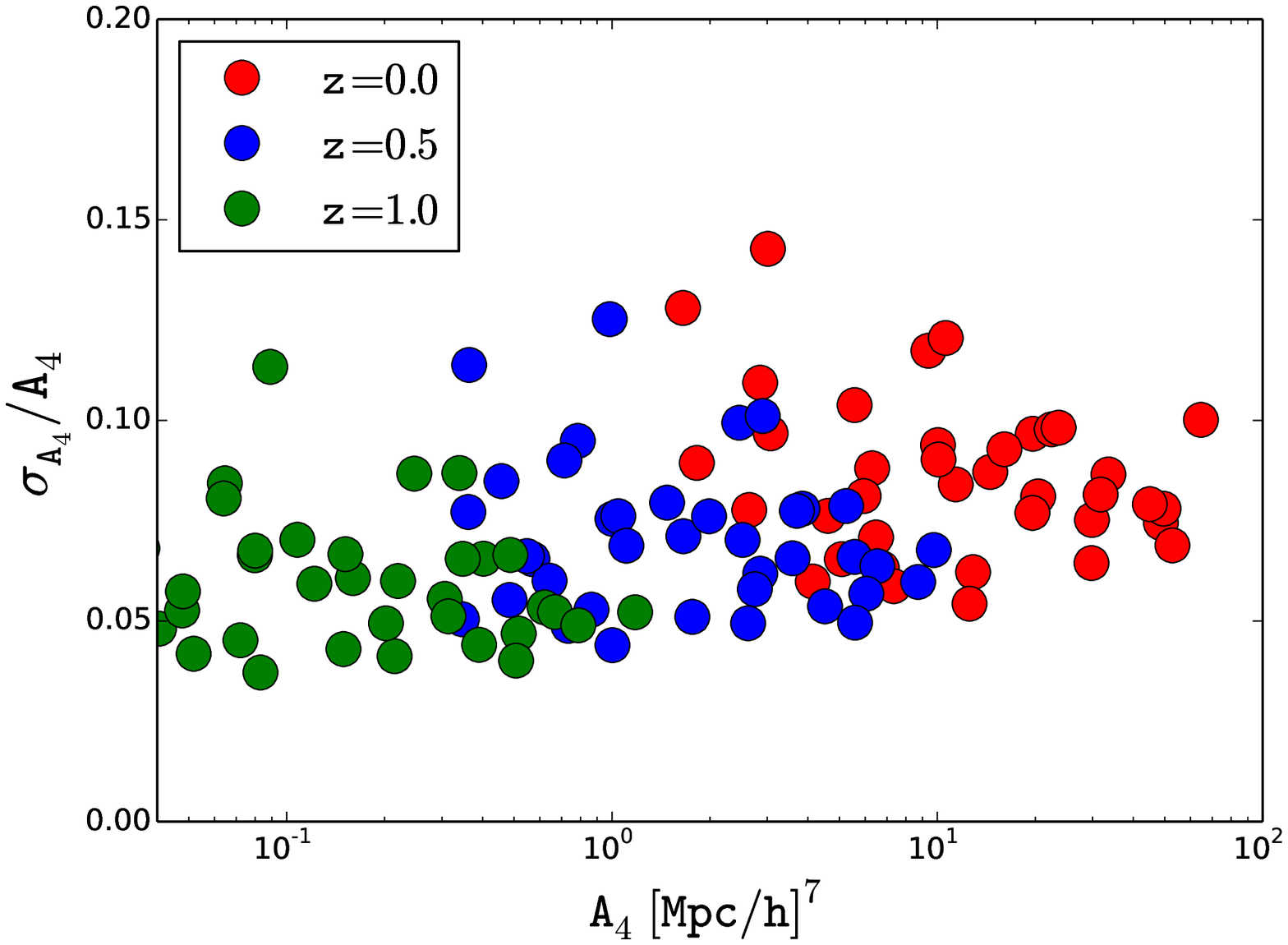}}
\caption{Relative variance $\Delta A_{2n}/A_{2n}$ versus $A_n$ based on our model for
$A_0$, $A_2$ and $A_4$ for three different redshift: 0.0 (red), 0.5 (blue) and 1.0 (green). Each circle bullet is one cosmological realization of the 38 cosmic emulator nodes.}
\label{fig:variance}
\end{figure*}

\begin{equation}
\begin{array}{lcl} 
A_0 & = & \int d\nu f(\nu) \dfrac{M}{\bar{\rho}}\ \Im_0\Im_0 \\ 
\\
A_2 & = & \int d\nu f(\nu) \dfrac{M}{\bar{\rho}}\ 2\Im_0\Im_1 \\ 
\\
A_4 & = & \int d\nu f(\nu) \dfrac{M}{\bar{\rho}}\ (\Im_1\Im_1 + 2\Im_0 \Im_2)
\end{array}
\label{A024}
\end{equation}
\\
with covariance,
\\
\begin{equation}
\mathtt{Cov}(A_iA_j) = 
 \begin{pmatrix}
  \int d\nu\ g(\nu)\ (\Im_0\Im_0)^2
  & \int d\nu\ g(\nu)\ (\Im_0\Im_0) (2\Im_0\Im_1)
  & \int d\nu\ g(\nu)\ (\Im_0\Im_0) (\Im_1\Im_1 + 2\Im_0 \Im_2)\\

  \int d\nu\ g(\nu)\ (\Im_0\Im_0) (2\Im_0\Im_1) 
  & \int d\nu\ g(\nu)\ (2\Im_0\Im_1)^2
  &  \int d\nu\ g(\nu)\ (2\Im_0\Im_1) (\Im_1\Im_1 + 2\Im_0 \Im_2)\\

  \int d\nu\ g(\nu)\ (\Im_0\Im_0) (\Im_1\Im_1 + 2\Im_0 \Im_2)
  & \int d\nu\ g(\nu)\ (2\Im_0\Im_1) (\Im_1\Im_1 + 2\Im_0 \Im_2)
  &  \int d\nu\ g(\nu)\ (\Im_1\Im_1 + 2\Im_0 \Im_2)^2
 \end{pmatrix}
 \label{eqn:covariance}
\end{equation}
where $i,j=0,2,4$ and
\\
\begin{equation}
	g(\nu) = \dfrac{1}{\mathtt{Volume}} f(\nu) \left(\dfrac{M}{\bar{\rho}}\  \right)^3
\end{equation}

In this paper we terminate this series after $A_4$ term. One can always go to higher order terms to get desired accuracy at higher $k$.

We will present the results of analytic calculations of $A_{2n}$ in the next section. 
Calculating the variance of each of these coefficients is as straightforward as calculating the coefficient itself, 
performing the integrals over the halo mass function. We calculate the variance on these terms for a volume of $1\ ({\rm 
Gpc}/h)^3$ for different cosmological models 
(the 38 models explained in next section) at three different redshifts: 0.0, 0.5 and 1.0. Figure \ref{fig:variance} shows the relative variance of the three coefficients. We find 
for 1\ ${\rm (Gpc)/h}^3$ the relative error $\sigma_{A_0}/A_0$ varies from 0.5 to 2 \%,
whereas on $\sigma_{A_2}/A_2$ and $\sigma_{A_4}/A_4$  vary from $1\%\ \rm{to}\ 7\%$ and from $2\%\ \rm{to}\ 20\%$, respectively, depending on the cosmology and redshift. 
We see that the relative error on $A_{2n}$ increases with $n$: this is a consequence of the fact that terms with higher 
$n$ receiving a larger contribution from higher mass objects, since the mass scaling of the 
integrand for $A_{2n}$ in the equations above is $M^{1+2n/3}$, while for the variance it is $M^{3+4n/3}$.
Higher mass objects are rarer and their Poisson fluctuations are larger, hence the relative 
variance is increased. 
Below we will compute the sensitivity of these parameters to cosmology: we will show that $A_0$ contains 
most of the information on the amplitude $\sigma_8$. 
In this paper we use halo mass function of \cite{2008ApJ...688..709T}. 

So far we assumed $F(k)=1$ without specifying its role. 
It was pointed out already in the original halo model \citep{2000MNRAS.318..203S} 
that the 1-halo term of the halo model fails to account for mass 
and momentum conservation at low $k$: the nonlinear corrections to the power spectrum have to scale as 
$k^4$ or $-k^2P(k)$ at low $k$, while the leading order of the 1-halo term scales as $k^0$. 
At very low $k$ such a term may even dominate over the linear term, which cannot be physical in 
the context of dark matter, even though it can happen in the context of galaxies \cite{2013PhRvD..88h3507B}. 
We will impose this constraint by simply fitting the residuals to the simulations at low $k$ and apply the derived transfer 
function $F(k)$, which vanished at low $k$,
to the model. We will show that the function $F(k)$ does not strongly depend on the cosmological 
model and we will thus ignore its dependence on cosmological parameters. 


\section{Calibrating the model with simulations}
\label{sec:calibration}

We use cosmic emulator (\cite{2010ApJ...715..104H,2009ApJ...705..156H,2010ApJ...713.1322L}) to evaluate the power spectra for 
each of the 38 emulator simulations and assume in each case it gives the true non-linear matter power spectrum. These reference power spectra are correct to nearly 1$\%$ up to $k\sim 1\ h \rm{Mpc}^{-1}$ at 38 different nodes (labelled as 0 to 37) in cosmological parameter space. This accuracy degrades to $5\%$ when computing the power spectrum away from the 
nodes.  Node 0 cosmology is closest to the WMAP-7 cosmology and we use it as a reference cosmology. 
We fit these simulation
power spectra with our description -- Quasi-linear Zeldovich term plus modified 1-halo term as a 
sum of even powers of $k$, to determine coefficients $A_{2n}$ as a function of cosmology.


\begin{figure*}
\centering
\subfigure{\includegraphics[width=0.45\textwidth]{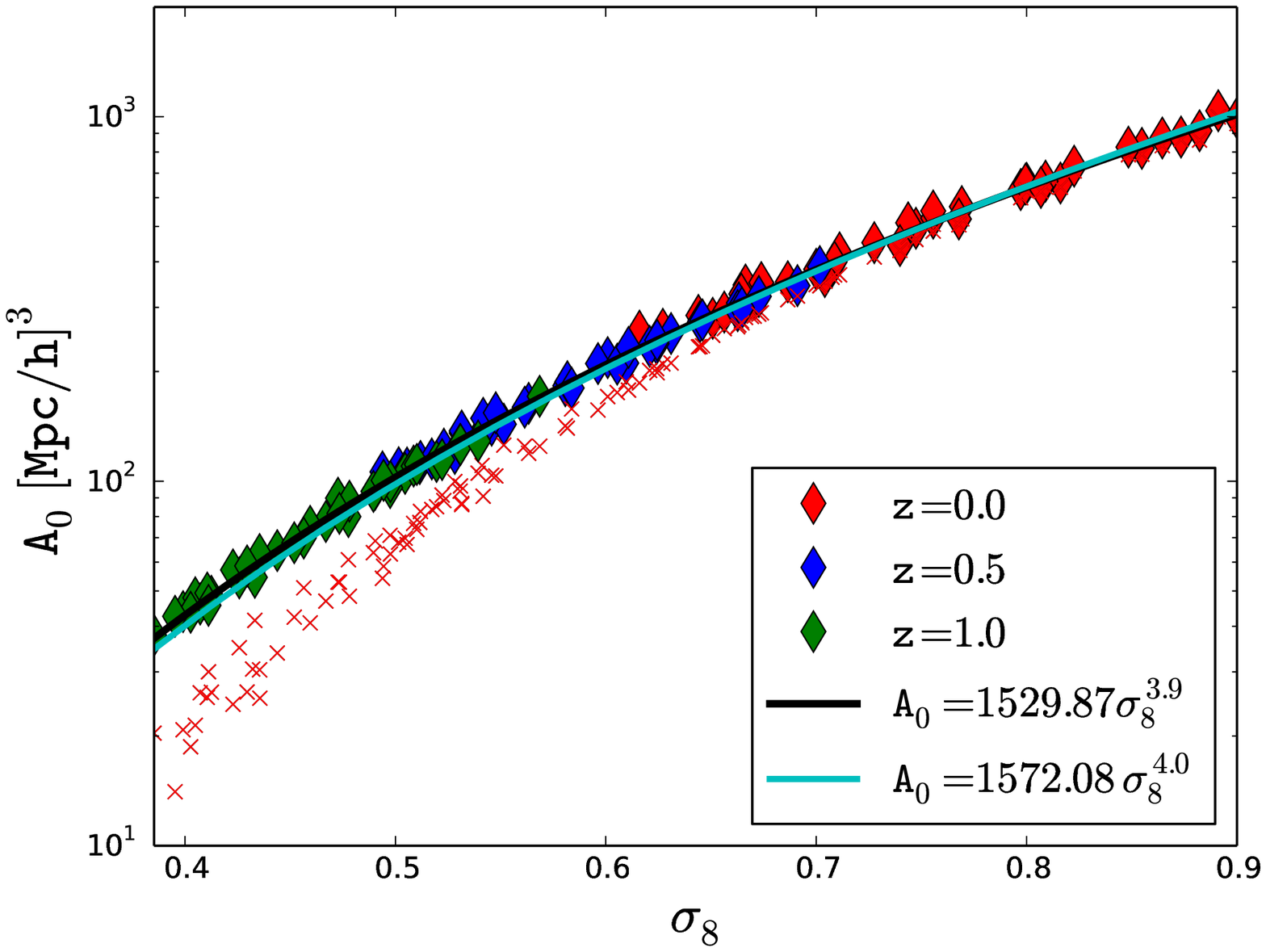}}
\subfigure{\includegraphics[width=0.45\textwidth]{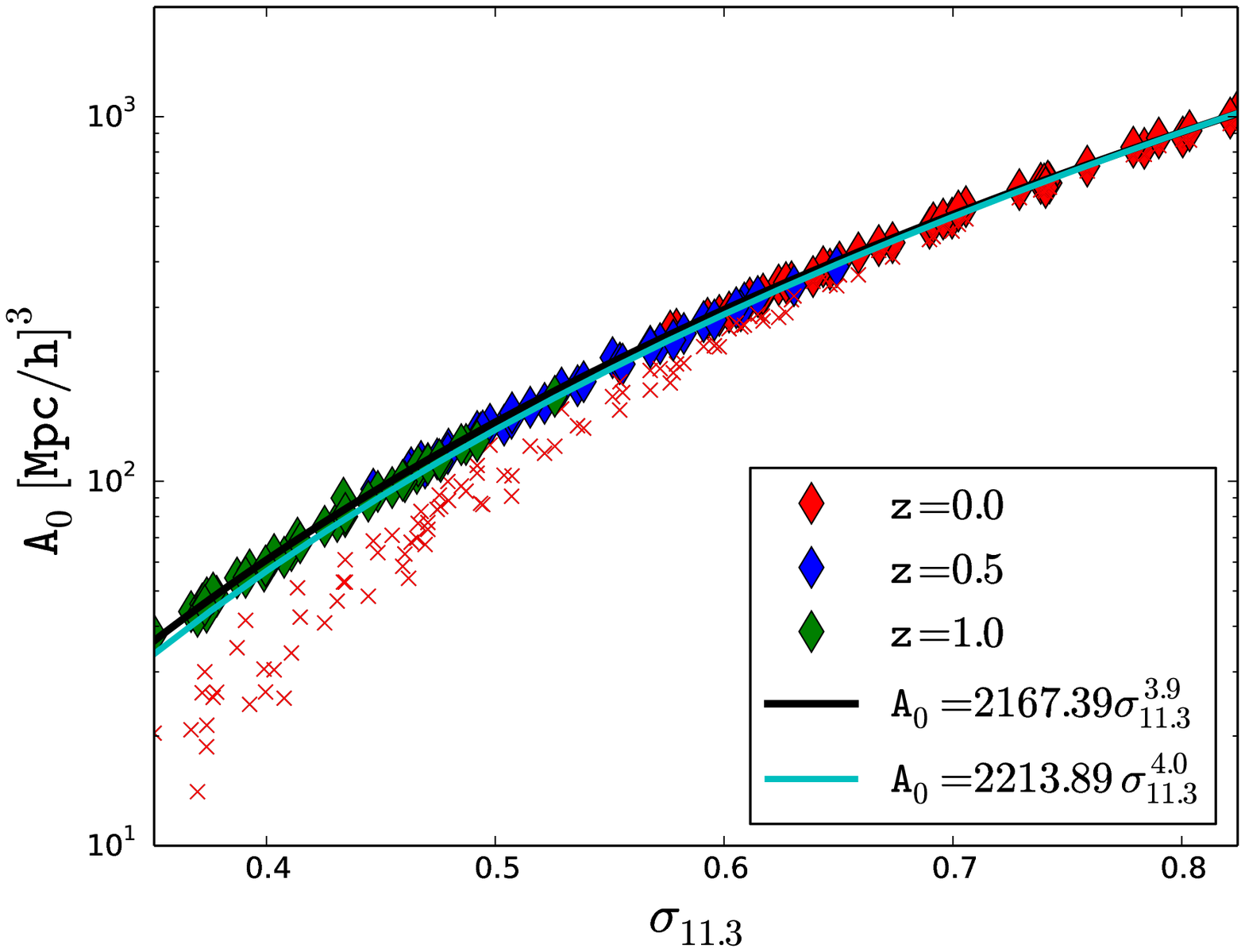}}
\subfigure{\includegraphics[width=0.45\textwidth]{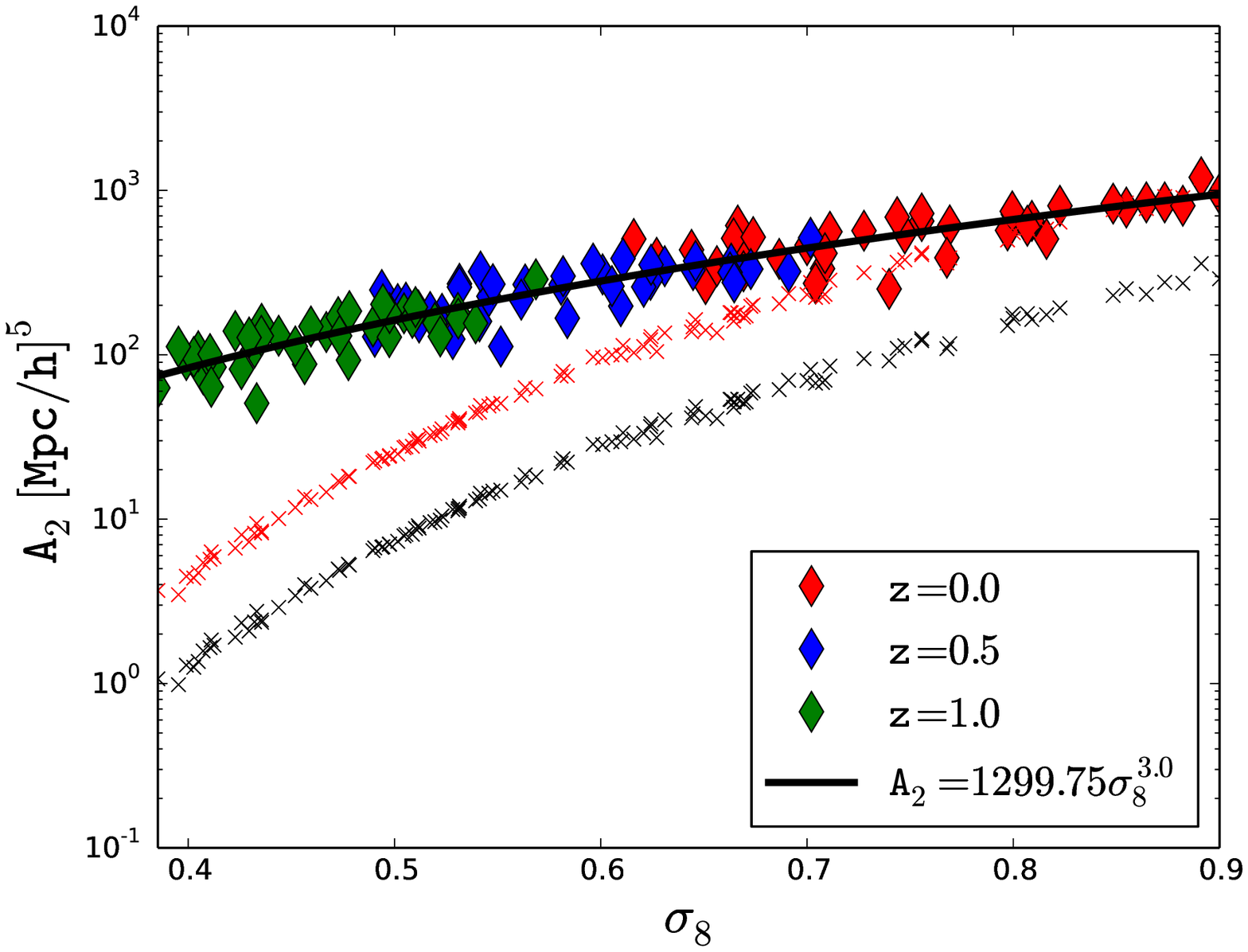}}
\subfigure{\includegraphics[width=0.45\textwidth]{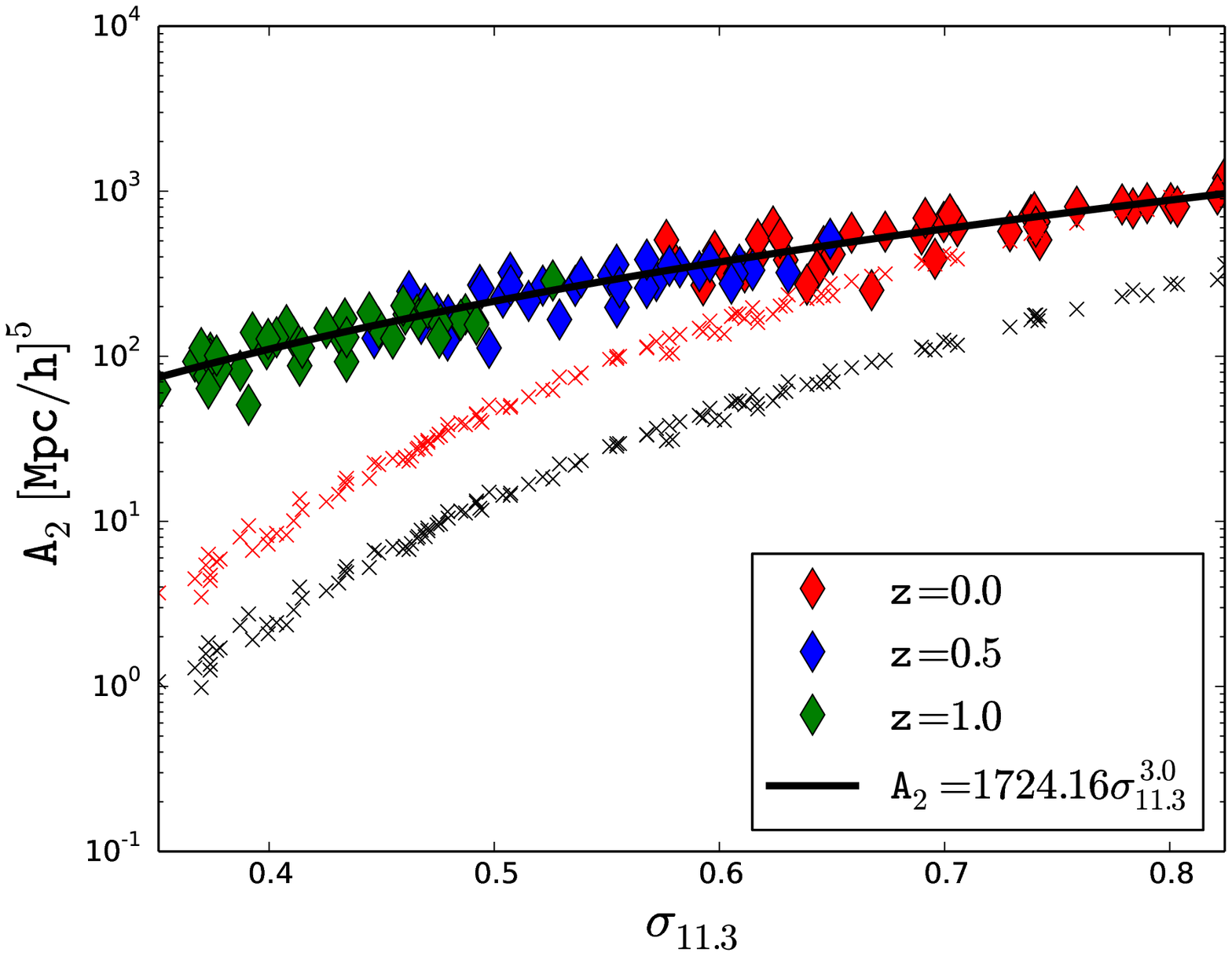}}
\subfigure{\includegraphics[width=0.45\textwidth]{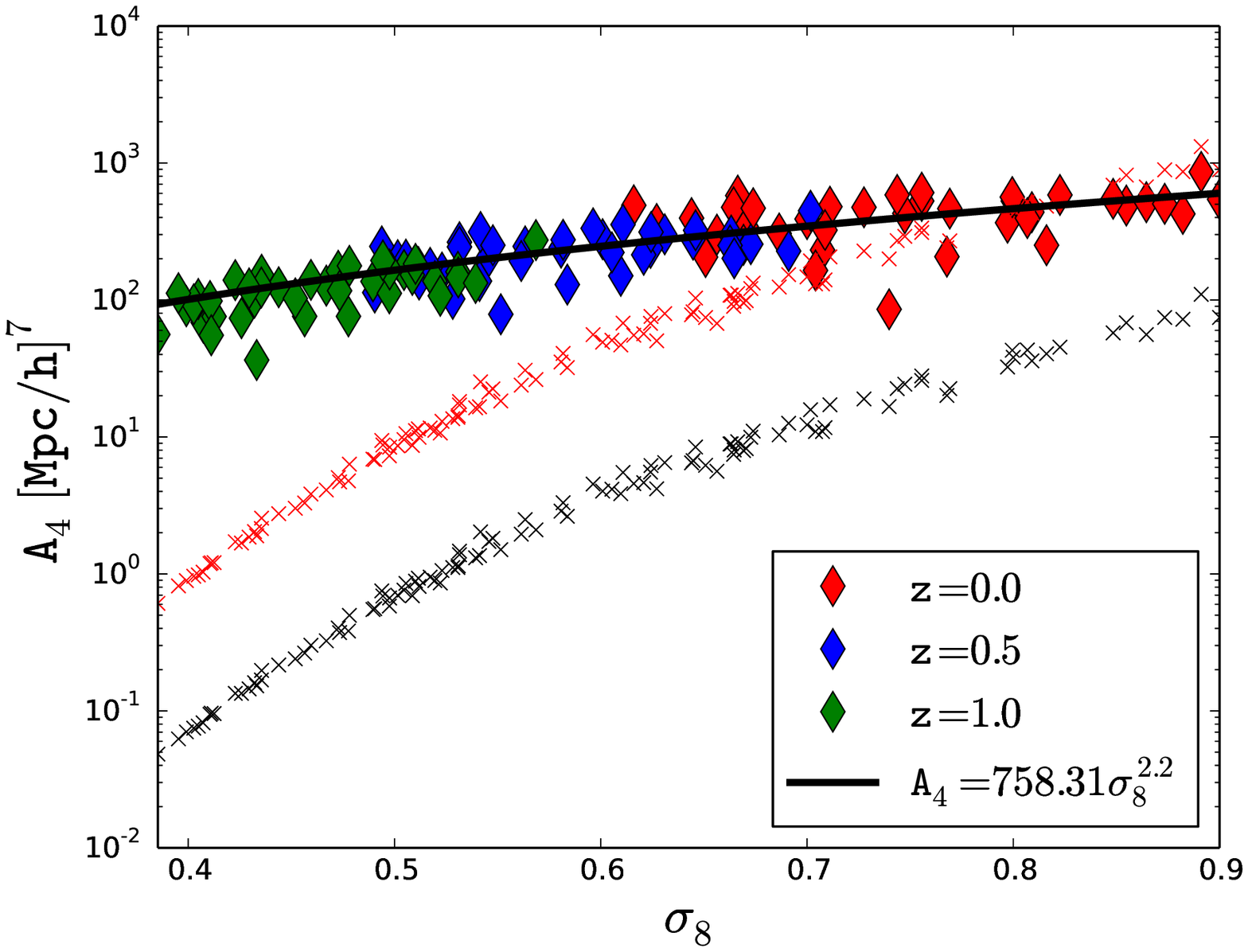}}
\subfigure{\includegraphics[width=0.45\textwidth]{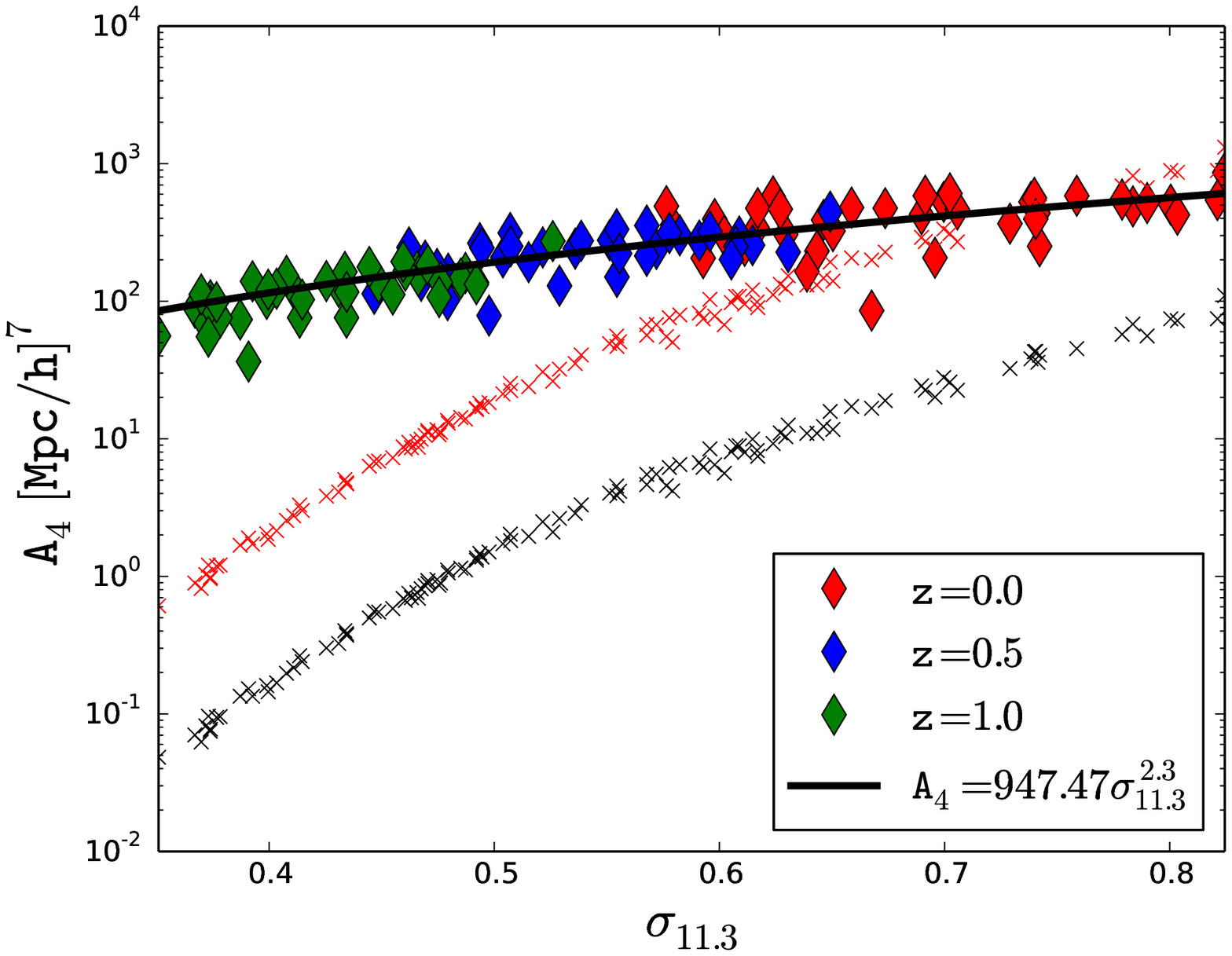}}
\caption{Fitted coefficients $A_0,\, A_2 \ \& \ A_4$ versus $\sigma_8$ (Left column) and $\sigma_{11.3}$ (Right column). 
We see that $\sigma_{11.3}$ reduces the scatter relative to $\sigma_8$ for $A_0$. Solid black line is the best fit power law stated in the legend. The halo model prediction is shown in crosses, using the usual halo concentration 
parameter $R_s=R_{\rm vir}/c$, with halos extending to the virial radius $R_{\rm vir}$, defined at the mean overdensity of 
200 (black crosses), and doubling that 
to $2R_{\rm vir}$ (red crosses). Halo model agrees well with simulations for $A_0$ 
at late redshifts, 
but not for $A_2$ and $A_4$ both in terms of amplitude and in terms of $\sigma_8$ or $\sigma_{11.3}$ scaling. Extending 
the halo profile to twice the virial radius improves the agreement.}
\label{fig:A_sigma}
\end{figure*}

To begin with, we fit the even power-law (equation \ref{eqn:pk1h_powerlaw} with $F(k)=1$) to the difference between matter power spectrum from emulator $\rm{P_{Emu}}$ and the Zeldovich term $\rm{P_{Zel}}$ for all 38 cosmologies and three redshifts: 0.0, 0.5, 1.0 between $k = 0.2$ and $0.8\ h \rm{Mpc}^{-1}$. 

All the coefficients fitted, $A_0, \, A_2 \ \& \ A_4$, are strongly correlated with $\sigma_8$, with 
$A_0$ having the least scatter. Figure \ref{fig:A_sigma} shows the scaling of these coefficients with $\sigma_8(z)$ and $\sigma_{11.3}$(z), where the latter was chosen to minimize the scatter in $A_0$.
Each of these coefficients can be approximately fit as a power law irrespective of the redshift and cosmology,
with $\sigma_8(z)$ scaling
\begin{equation}
 	A_0 \propto \sigma_{8}^{3.9},\ A_2 \propto \sigma_{8}^{3.0},\ A_0 \propto \sigma_{8}^{2.2}.
\label{slopea024}
\end{equation}
It is not straightforward to determine the errors since this is not a formal fit to a set of data points with individual 
errors. In figure \ref{fig:A_sigma} we also show results when the slope of $A_0$ is 4.0: we see this is also a good fit over 
the range. 

Figure \ref{fig:A_sigma} also shows the predictions of the halo model for these coefficients (in black crosses). While the 
halo mode predicts well $A_0$ at low redshifts, it fails for higher order coefficients. This can be improved 
if the virial radius is increased by roughly a factor of 2 at low redshifts, and more than that at higher redshifts
(which needs to be taken to power $2n$ to evaluate the effect on $A_{2n}$), shown as red crosses in figure \ref{fig:A_sigma}. The failure of the halo model to quantitatively 
predict these coefficients is not surprising: the halos do not suddenly stop at the virial radius and the halo model has 
some flexibility in how it is implemented. Our goal here is not to understand the halo model, but to have accurate 
predictions.  For this reason we will just use the fits of $A_{2n}$ coefficients to simulations in this paper. 

\begin{figure*}
\centering
\subfigure{\includegraphics[width=0.48\textwidth]{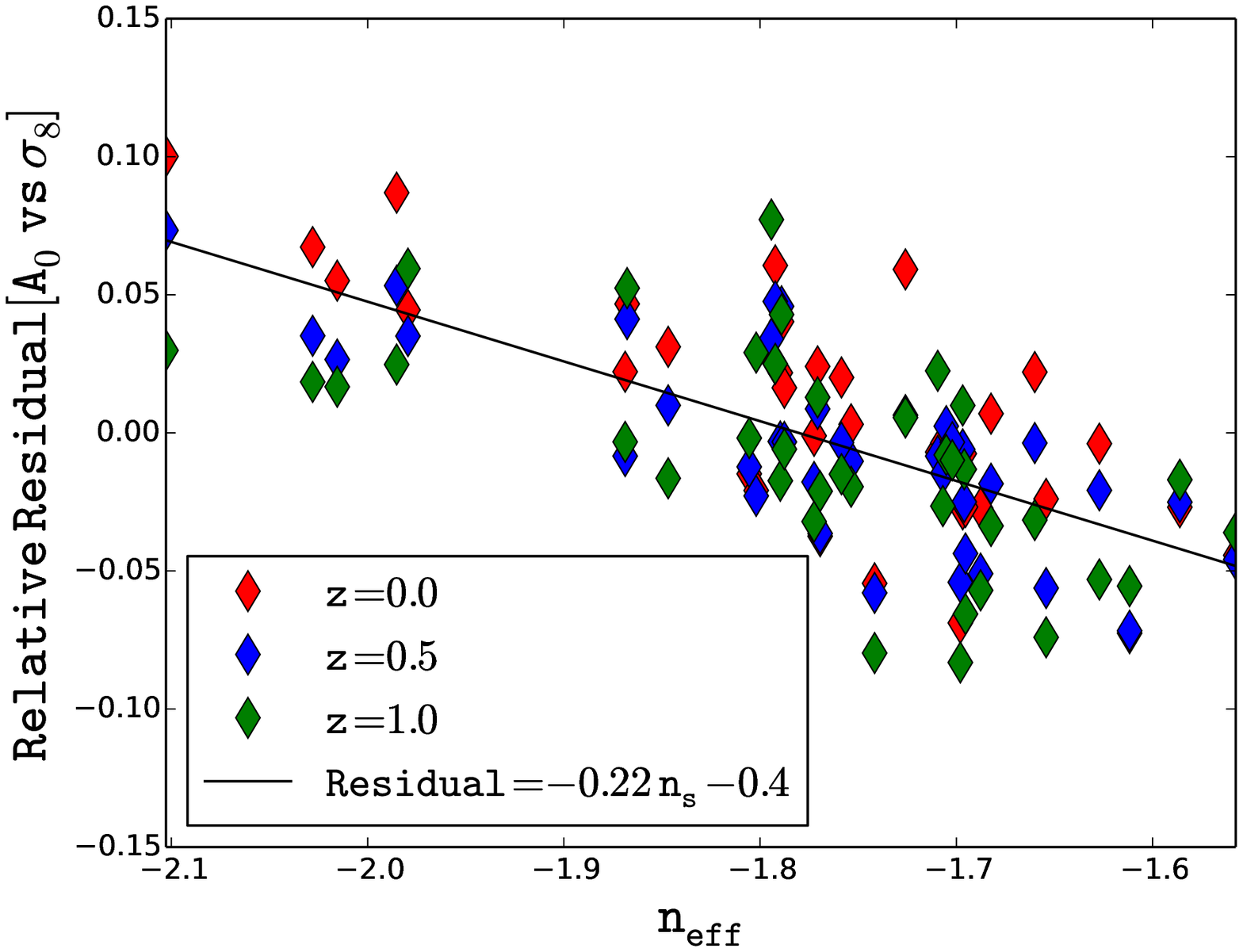}}
\subfigure{\includegraphics[width=0.48\textwidth]{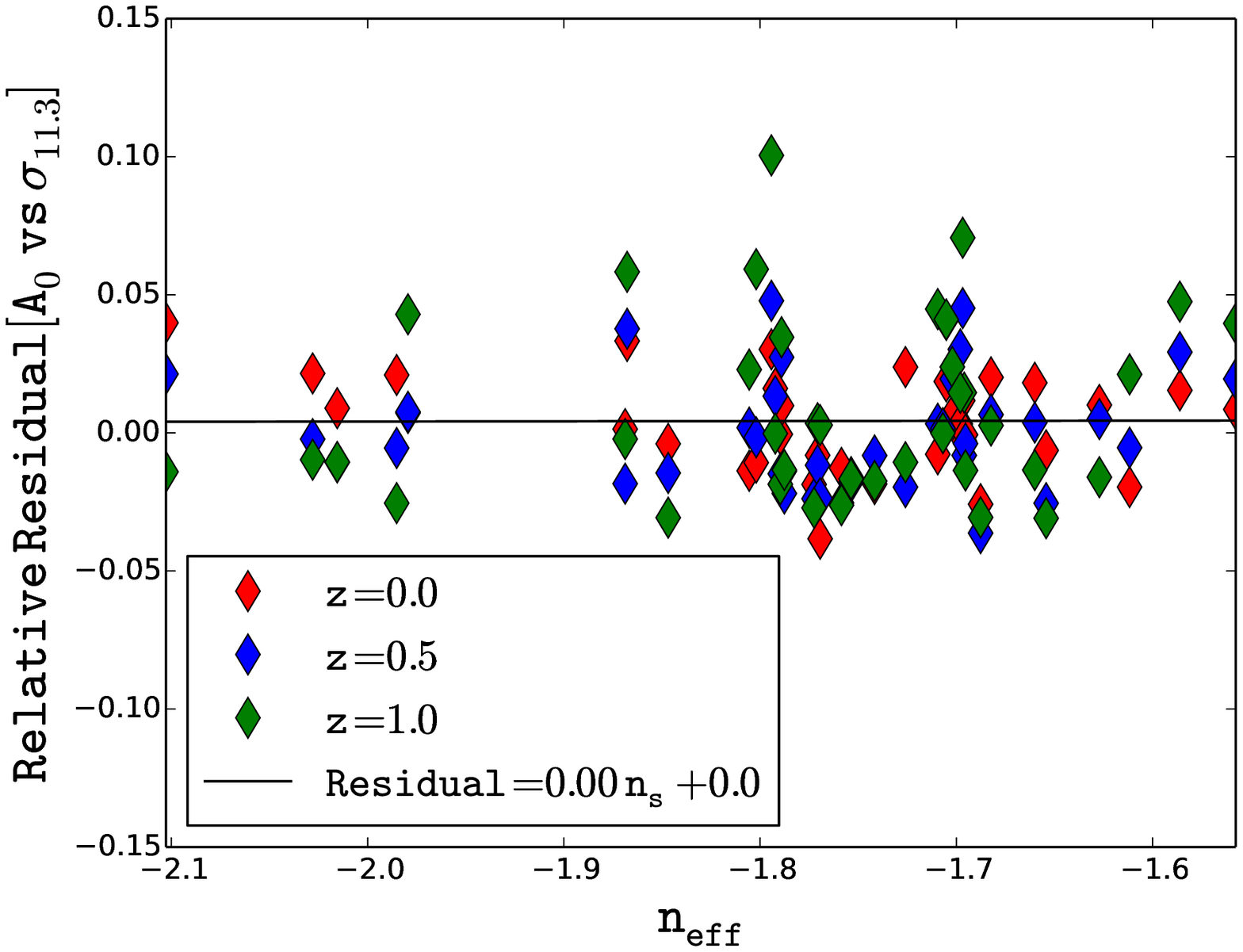}}
\subfigure{\includegraphics[width=0.48\textwidth]{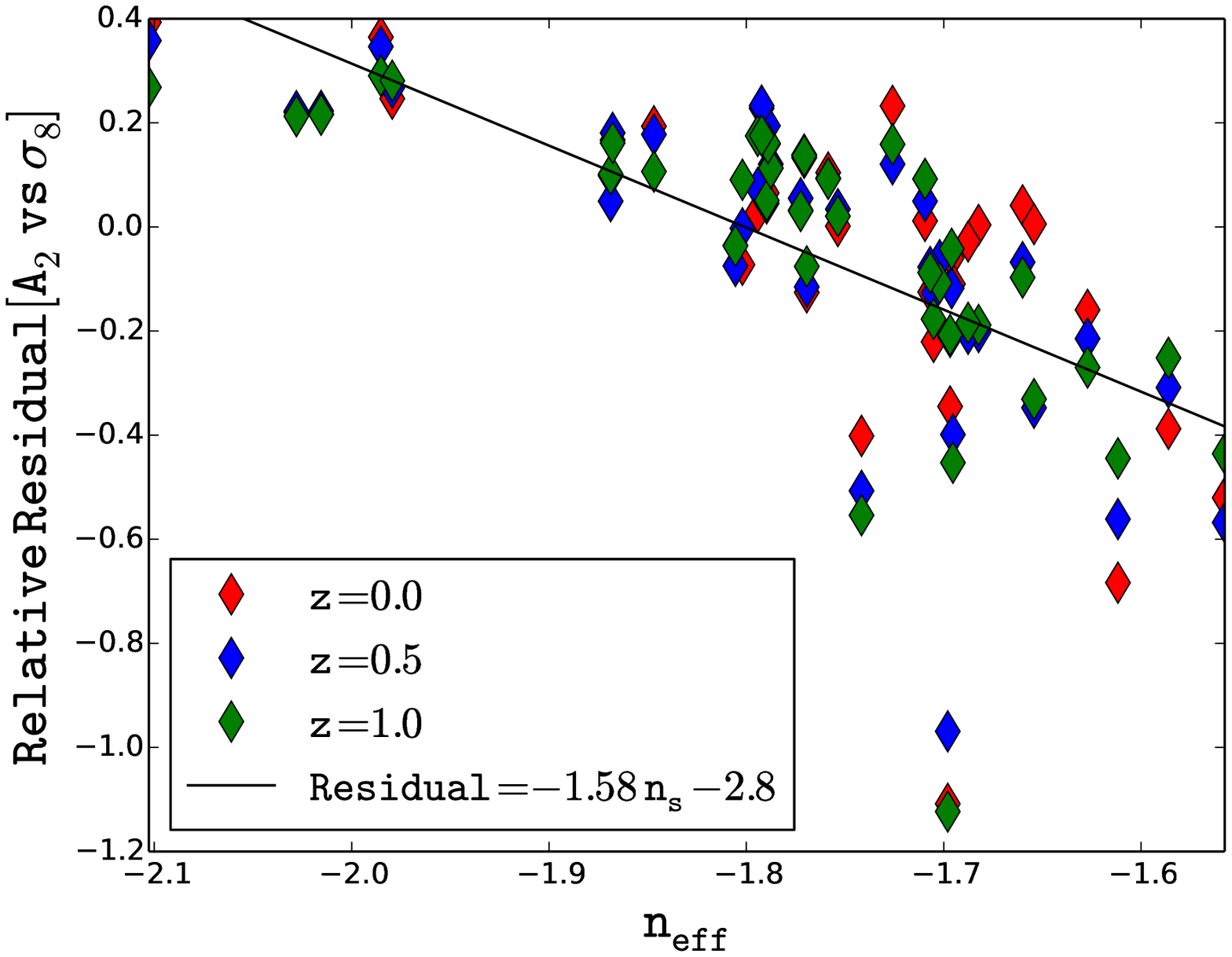}}
\subfigure{\includegraphics[width=0.48\textwidth]{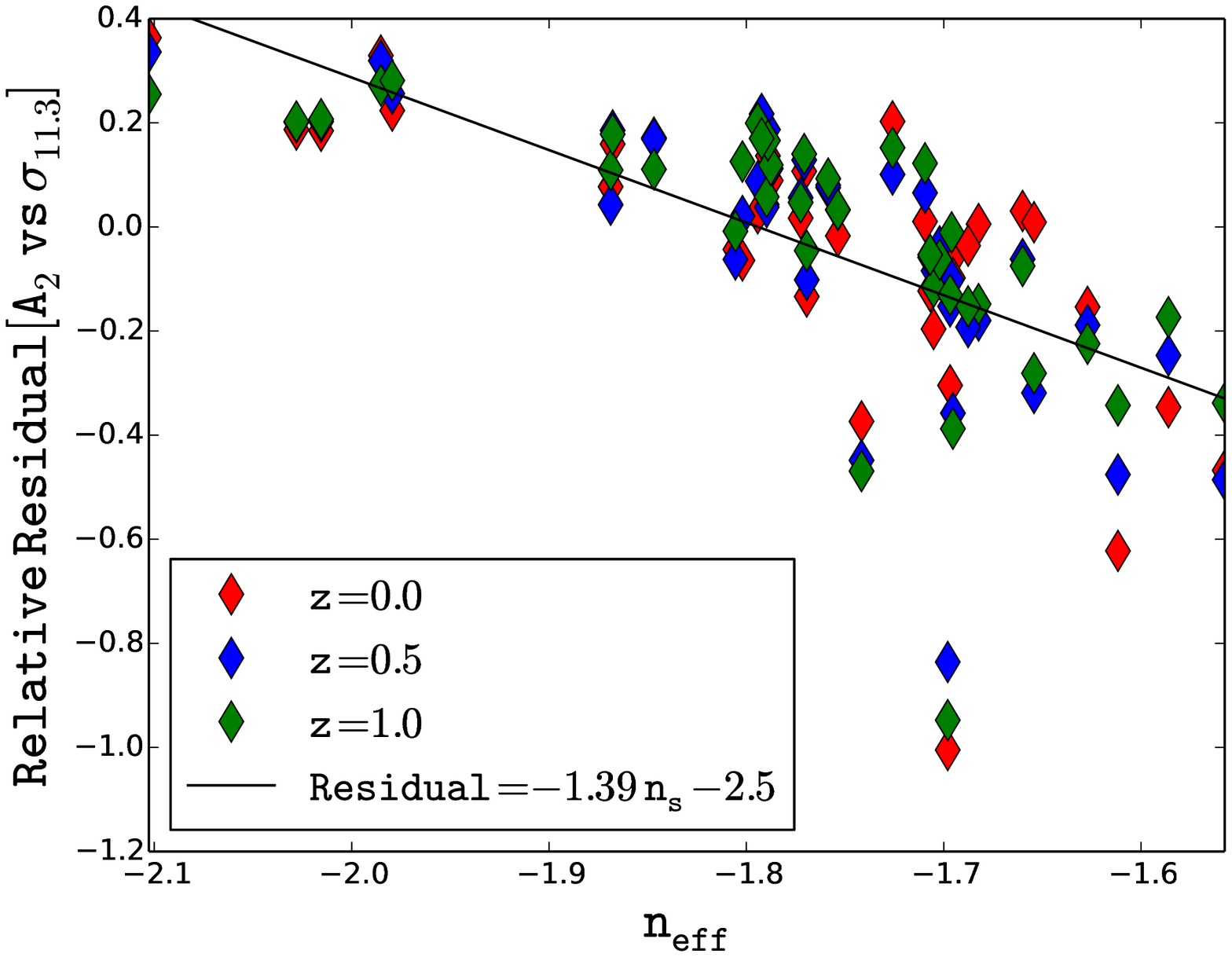}}
\subfigure{\includegraphics[width=0.48\textwidth]{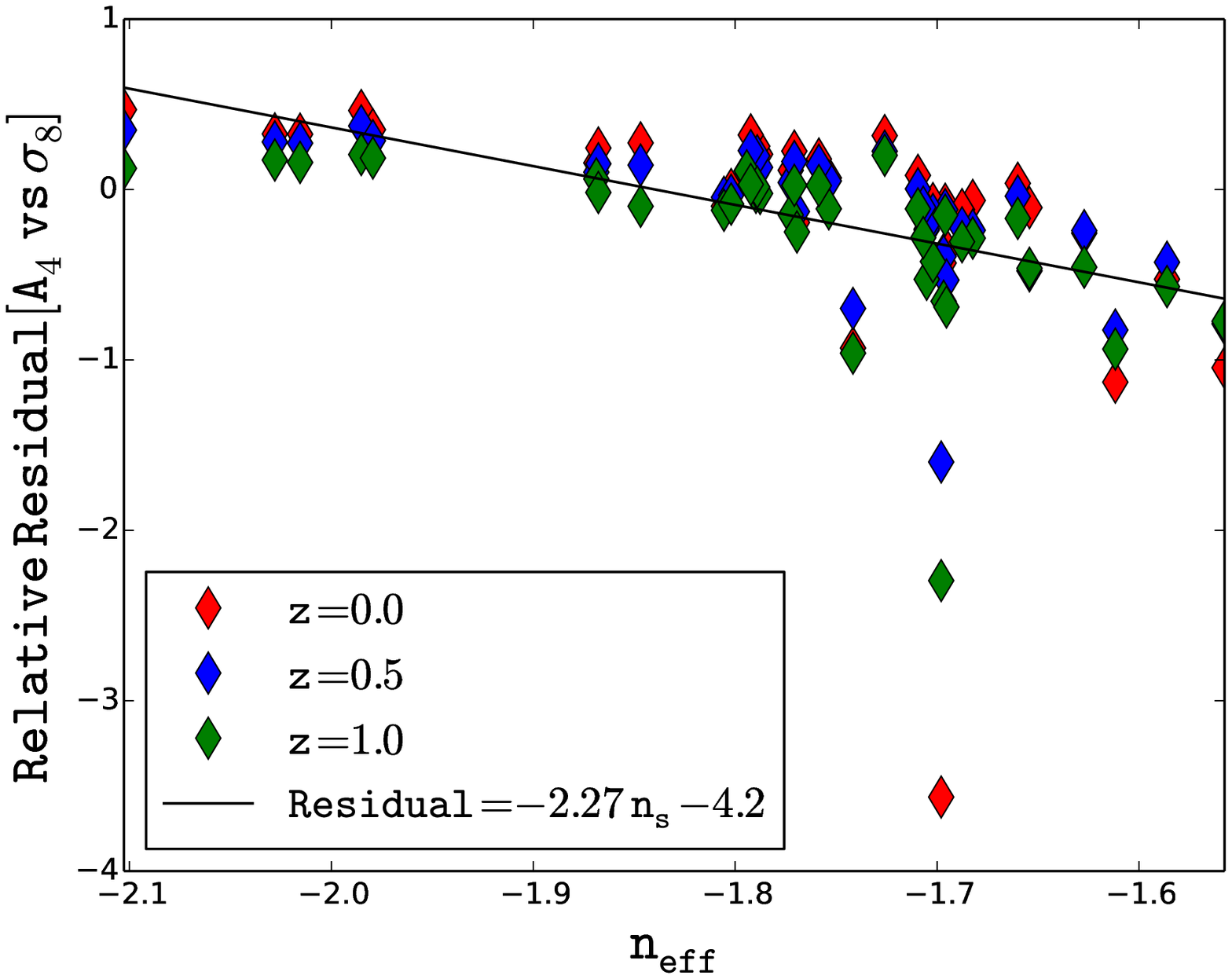}}
\subfigure{\includegraphics[width=0.48\textwidth]{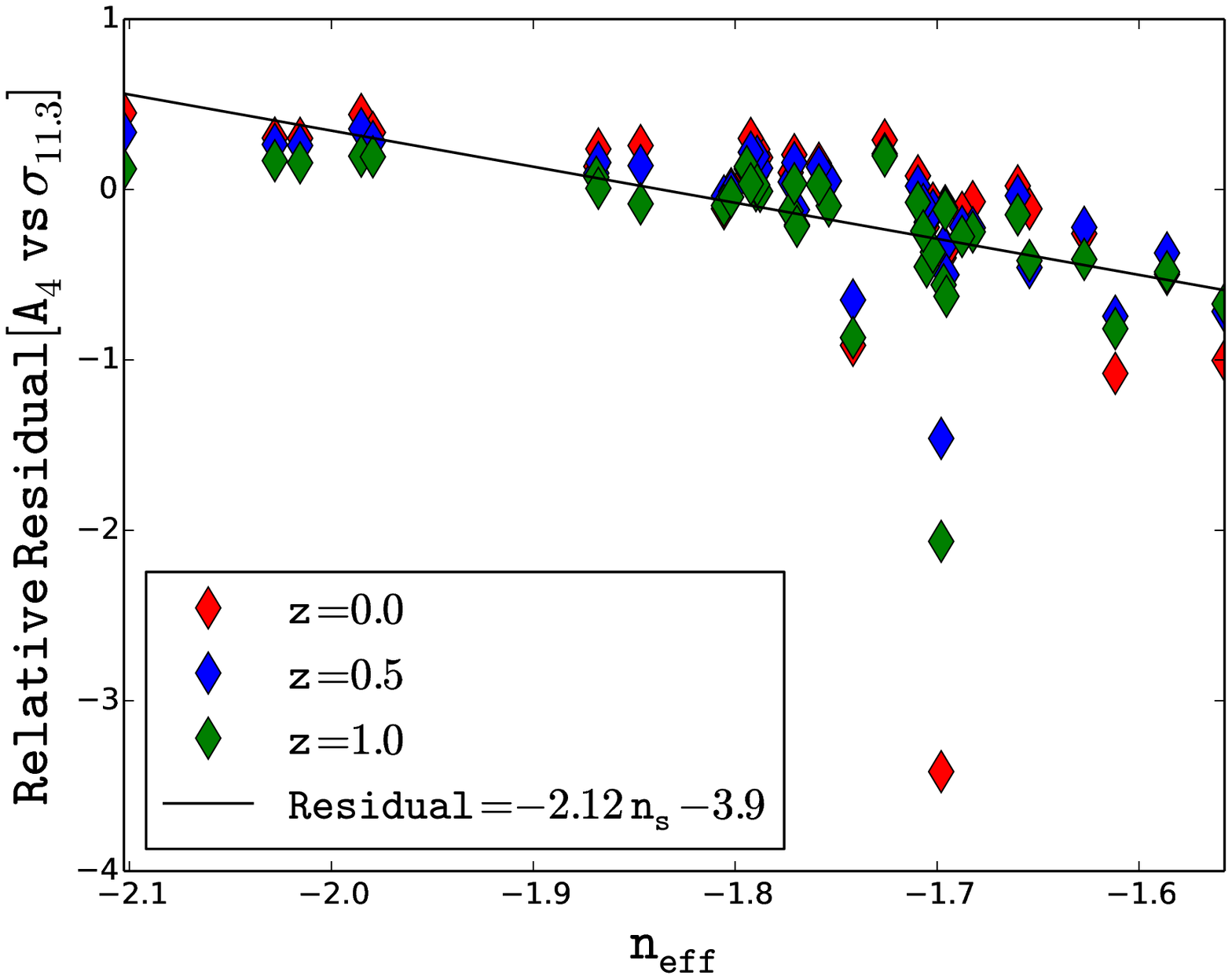}}

\caption{Correlation between effective slope ($n_{\rm eff}$) and 
residuals after $\sigma_8$ (Left column) or $\sigma_{11.3}$ (Right column) scaling is taken out 
and their respective best fit. Solid black line is the best linear fit as stated in the legend.}
\label{fig:ns_res}
\end{figure*}

The next step is to correct for the scatter around the best fit $\sigma_{8}$. A correlation is noticed between the residual of the coefficients with the effective slope $n_{\rm eff}$. This is shown in figure \ref{fig:ns_res}. Here the residual means the difference between the diamond-bullets and best fit lines in figure \ref{fig:A_sigma} and 
effective slope $n_{\rm eff}$ 
is calculated as the slope of the linear matter power spectrum at $k \sim 0.2\ h \rm{Mpc}^{-1}$. The higher order coefficients have larger 
scatter and stronger correlation between this residual and effective slope. We tested the scalings for 
few different values of $R$ in $\sigma_R$ and found minimum scatter for $\sigma_{11.3}$, 
which can be seen in figures \ref{fig:A_sigma} and \ref{fig:ns_res}. By using $\sigma_{11.3}$ instead of $\sigma_8$ one can remove the correlation with effective slope for $A_0$, so no $n_{\rm eff}$ correction is needed for $A_0$. However, $A_2 \ \& \ A_4$ 
still need to be corrected for this correlation, although
the correction is smaller in case of $\sigma_{11.3}$ than $\sigma_8$. Hence the corrected expressions for these coefficients are
\begin{equation}
	A_0 = 1529.87 \sigma_8^{3.9} \times (1 + [-0.22 n_{\rm{eff}} - 0.4]),\ \rm{or}\ 
	A_0 = 2167.39 \sigma_{11.3}^{3.9},
	\label{eqn:a0}
\end{equation}
\begin{equation}
	A_2 = 1299.75 \sigma_8^{3.0} \times (1 + [-1.58 n_{\rm{eff}} - 2.8]),\ \rm{or}\ 
	A_2 = 1724.16 \sigma_{11.3}^{3.0} \times (1 + [-1.39 n_{\rm{eff}} - 2.5])
	\label{eqn:a2}
\end{equation}
\begin{equation}
	A_4 = 758.31 \sigma_8^{2.2} \times (1 + [-2.27 n_{\rm{eff}} - 4.2]),\ \rm{or}\ 
	A_4 = 947.47 \sigma_{11.3}^{2.3} \times (1 + [-2.12 n_{\rm{eff}} - 3.9])
	\label{eqn:a4}
\end{equation}

\begin{figure*}
\centering
\subfigure{\includegraphics[width=0.48\textwidth]{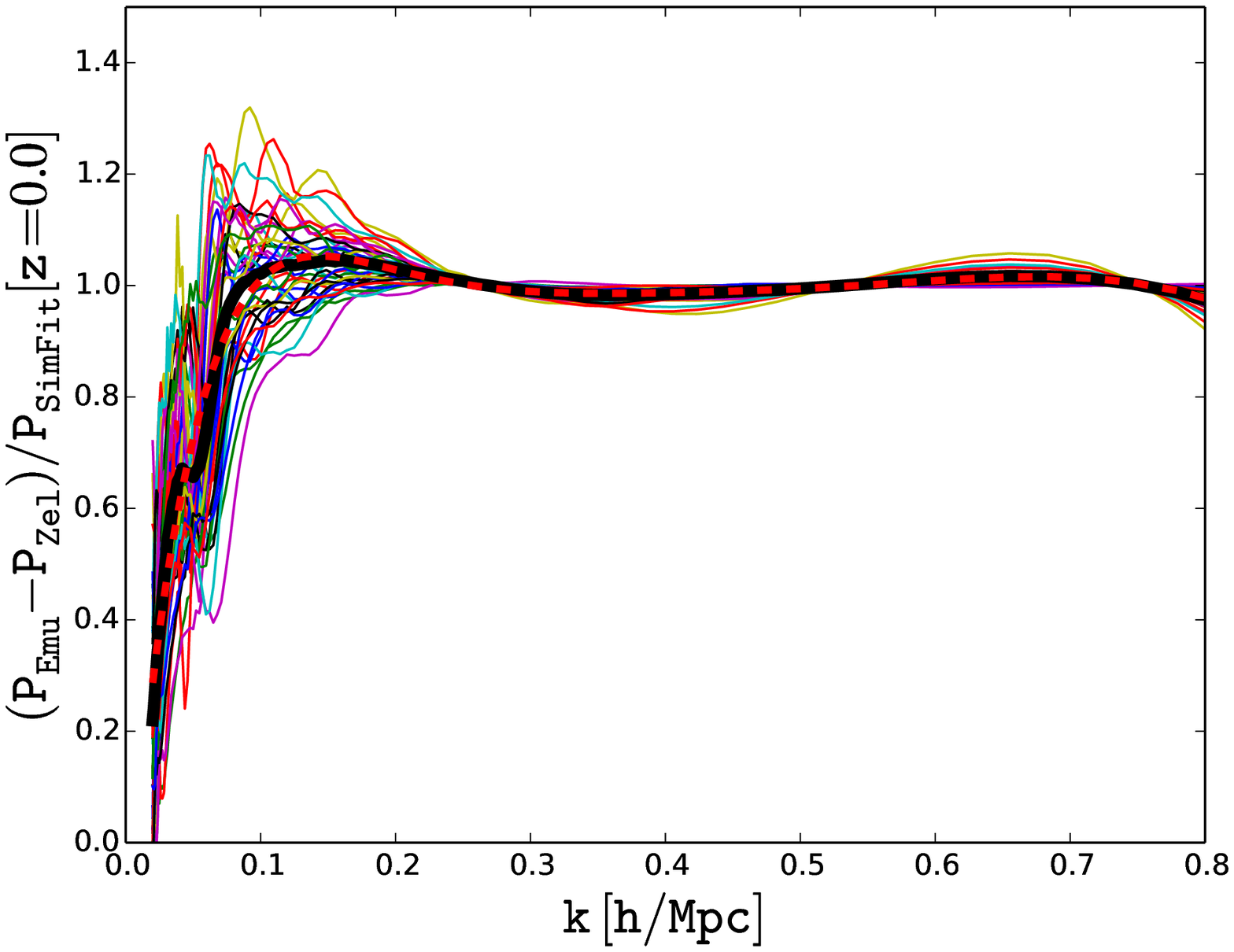}}
\subfigure{\includegraphics[width=0.48\textwidth]{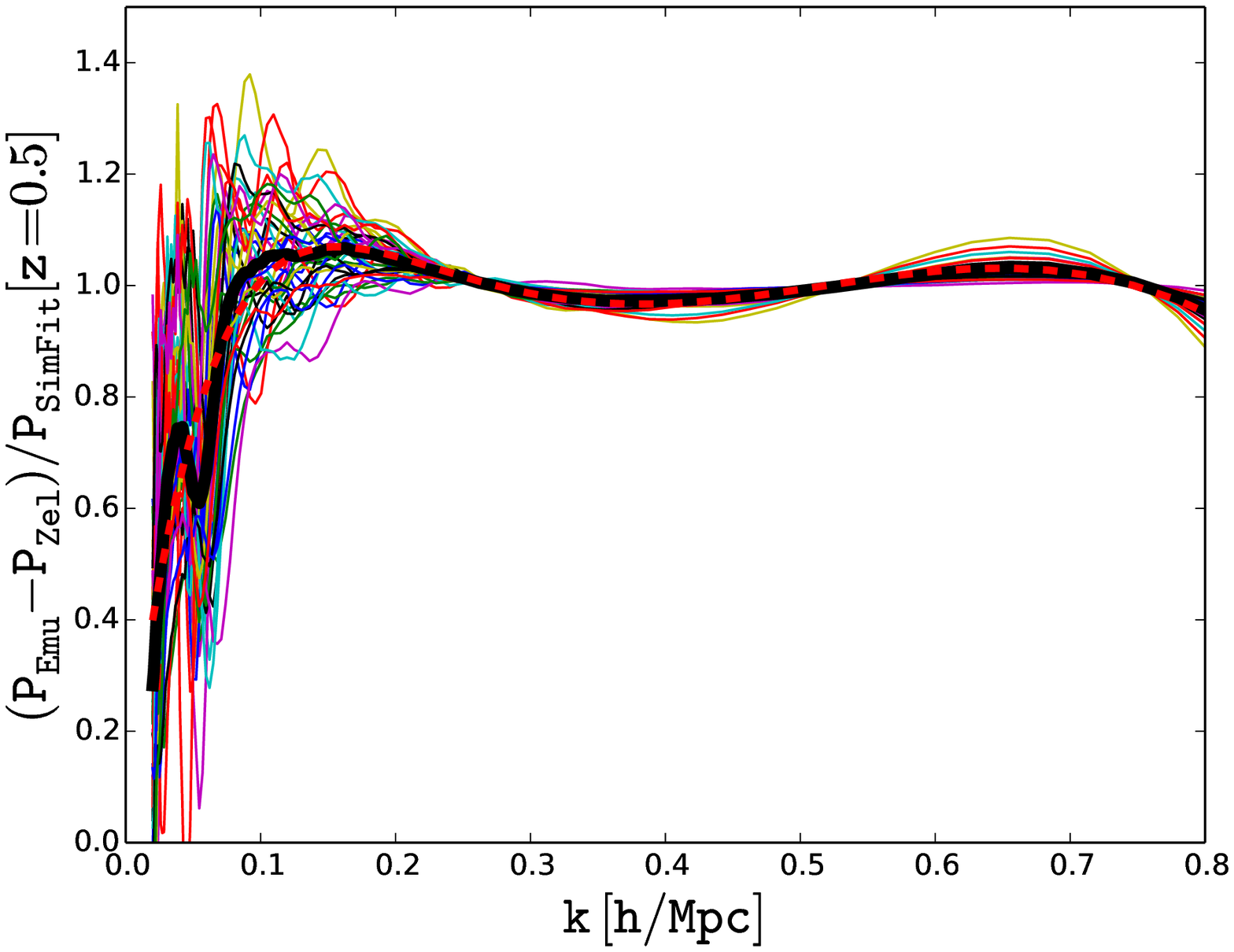}}
\subfigure{\includegraphics[width=0.48\textwidth]{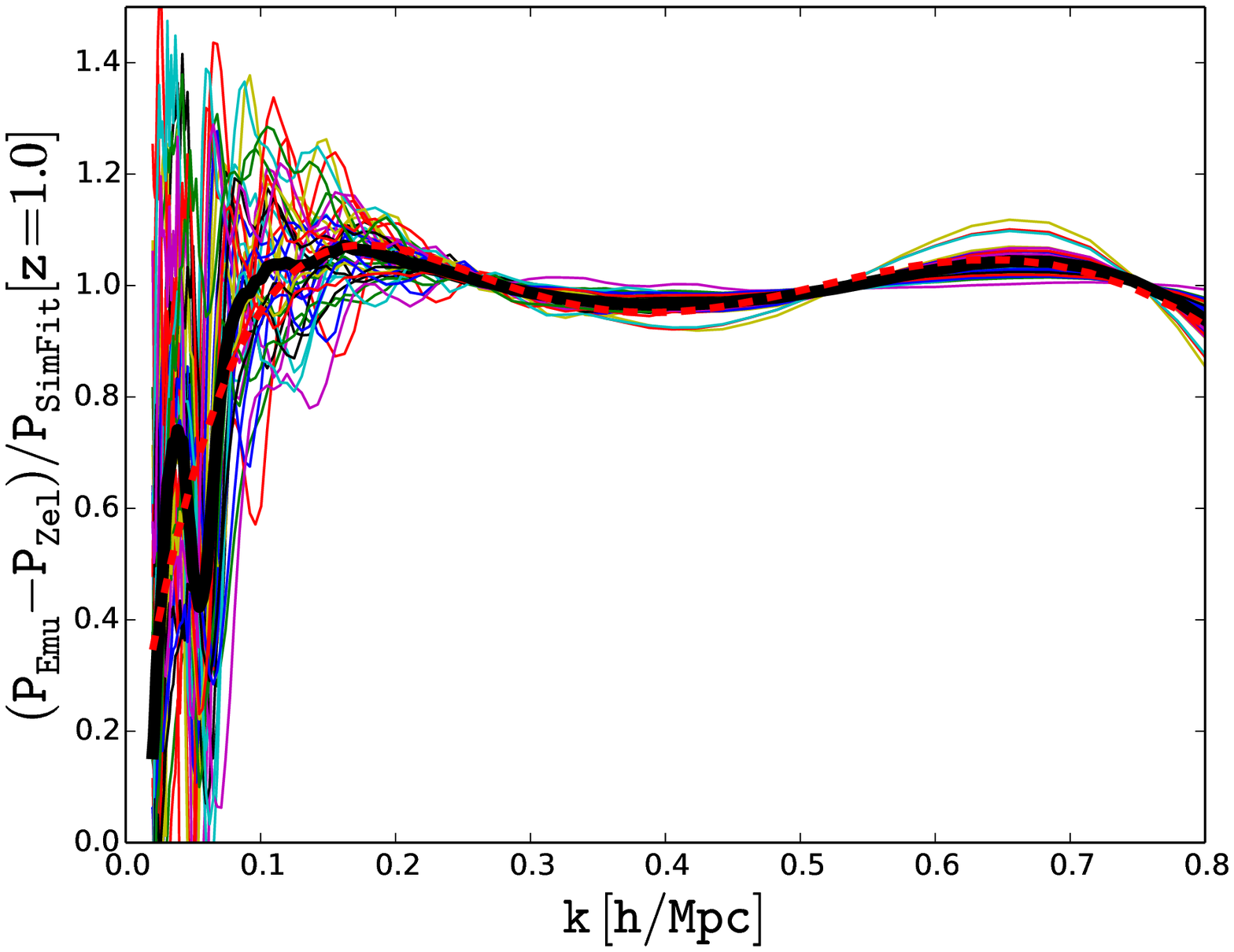}}
\subfigure{\includegraphics[width=0.48\textwidth]{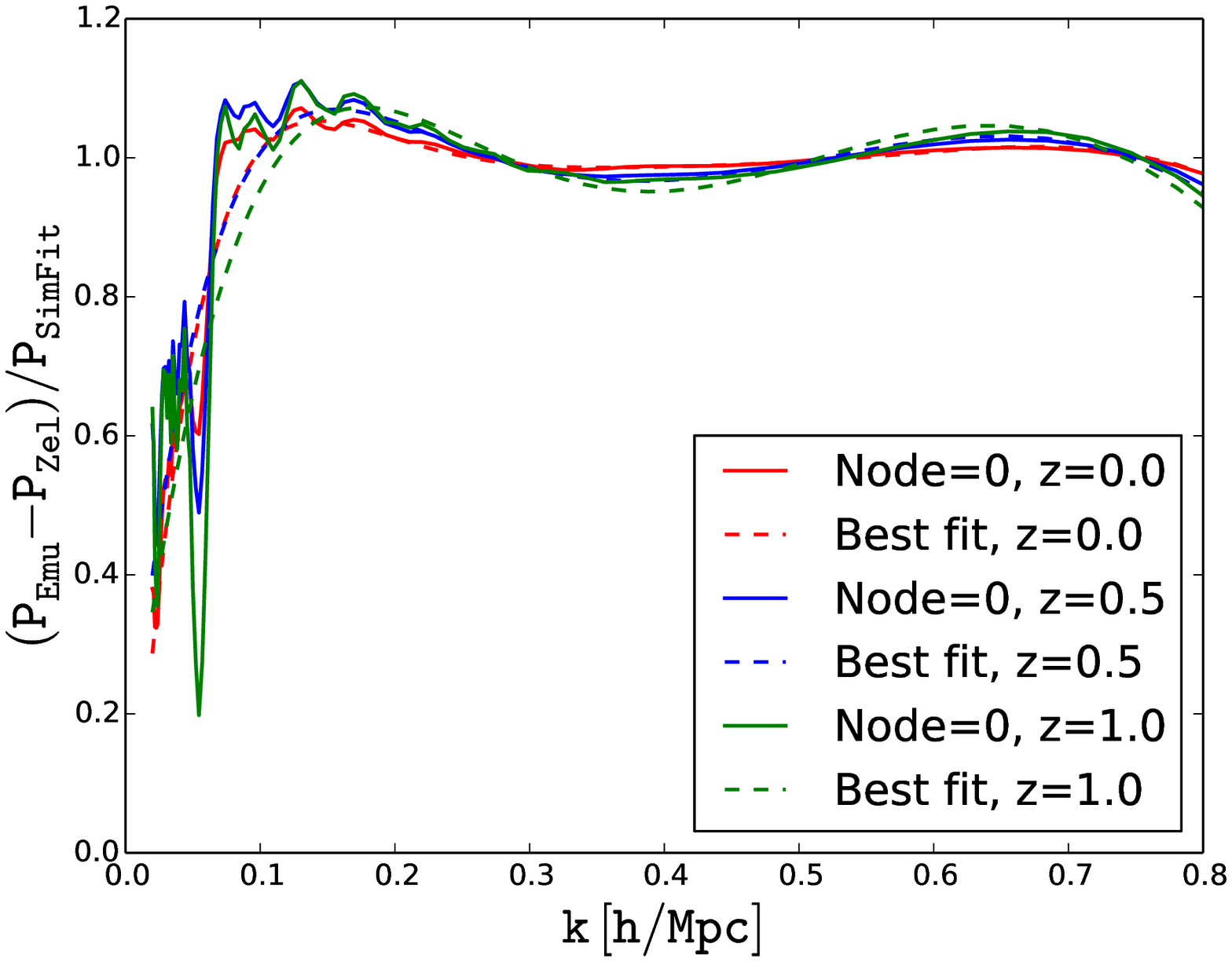}}

\caption{The first three panel (in reading order), shows the ratio of $(\rm{P_{emu} - P_{Zel}})$ and $\rm{P_{SimFit}} = A_0-A_2k^2+A_4k^4$, where the coefficients $A_0, A_2, A_4$ are the best fit coefficients to the emulator matter power spectrum for all 38 cosmological models (in different colors) at three different redshifts: 0.0 (top-left), 0.5 (top-right) and 1.0 (bottom-left). Bottom-right panel shows same quantity for node 0 and the best fit to the average (of 38 coloured curves in first three panels) at three different redshifts.}
\label{fig:bestfit}
\end{figure*}

We still need to account for the mass conservation, which forces the 1-halo term to go to 0 at low $k$. 
In figure \ref{fig:bestfit}, we plot the ratio of the difference between $\rm{P_{Emu}}$ and $\rm{P_{Zel}}$ with $\rm{P_{SimFit}}$ which is given by
	$\rm{P_{SimFit}} = A_0 - A_2k^2 + A_4 k^4$, 
where, these coefficients are the best fit values to $(\rm{P_{emu} - P_{Zel}})$ for all 38 cosmologies (Diamond bullets in figure \ref{fig:A_sigma}). The top-left, top-right and bottom-left panel shows the same quantity at three different redshifts: 0.0, 0.5 and 1.0 respectively. All 38 curves in each panel are very close to 1 for $k$ between 0.2 and $0.8 \ h \rm{Mpc}^{-1}$, which is expected as these coefficients are fitted in that range in the first place. Outside this range the scatter increases. We took the average of all these 38 curves at all three redshifts and fit it to a $10^{th}$ 
order polynomial, requiring to vanish at low $k$. 
The bold solid black line and dashed red curve represents the average and best fit to the average, respectively. 
It can be seen in bottom-right panel of figure \ref{fig:bestfit} that these best fit to the average are very close to node 0 cosmology curve and also very close to each other for different redshifts for $k<0.8 \ h \rm{Mpc}^{-1}$. 
We average of these three best-fit curves, at three different redshifts, to build the function $F(k)$ for the 1 halo term, 
which we model as
\begin{equation}
	F(k) = \sum_{n=0}^{10} a_n k^n, 
\label{eqn:fk}
\end{equation}
where the coefficients $a_n$ are listed in table \ref{tbl:an}. 
As expected by the mass conservation arguments, and seen in figure \ref{fig:bestfit},
this correction drops to zero for $k<0.1\ h \rm{Mpc}^{-1}$. 
In principle we should force it to go to 0 as $k^2$, but we found this caused problems to 
the fit at higher $k$: the effects of $F(k)$ are very small in any case and in most instances below 1\%, since
at low $k$ the Zeldovich term dominates. 
For this reason we will assume this correction is 
independent of the cosmological model or redshift. 

\begin{table*}
\begin{center}
\begin{tabular}{|cccccccccccc|}
\hline
\hline
 $a_n$ & $a_0$ & $a_1$& $a_2$& $a_3$& $a_4$& $a_5$& $a_6$& $a_7$& $a_8$& $a_9$& $a_{10}$ \\
\hline
value &0.0 & 21.814& -174.134& 747.369 & -2006.792 & 3588.808 & -4316.241 & 3415.525 & -1692.839 & 474.377 & -57.228\\
\hline
\hline
\end{tabular}
\caption{Coefficients to calculate the correction function, equation \ref{eqn:fk}.
The units of the coefficient $a_n$ is $(\rm{Mpc}/h)^n$.}
\label{tbl:an}
\end{center}
\end{table*}

We combine the above two terms to obtain the matter power spectrum as:

\begin{equation}
	P(k,z) = P_{\rm zel}(k,z) + P_{\rm 1h}(k,z)
	\label{eqn:final}
\end{equation}
\\
and,
\begin{equation}
	P_{\rm 1h}(k,z) = (A_0 - A_2k^2 + A_4k^4)F(k)
\end{equation}
\\
where, $A_0, A_2 \ \& \ A_4$ are given by equation \ref{eqn:a0}, \ref{eqn:a2} and \ref{eqn:a4}, respectively,
and $F(k)$ is given by equation \ref{eqn:fk}. 

\begin{figure*}
\centering
\subfigure{\includegraphics[width=0.48\textwidth]{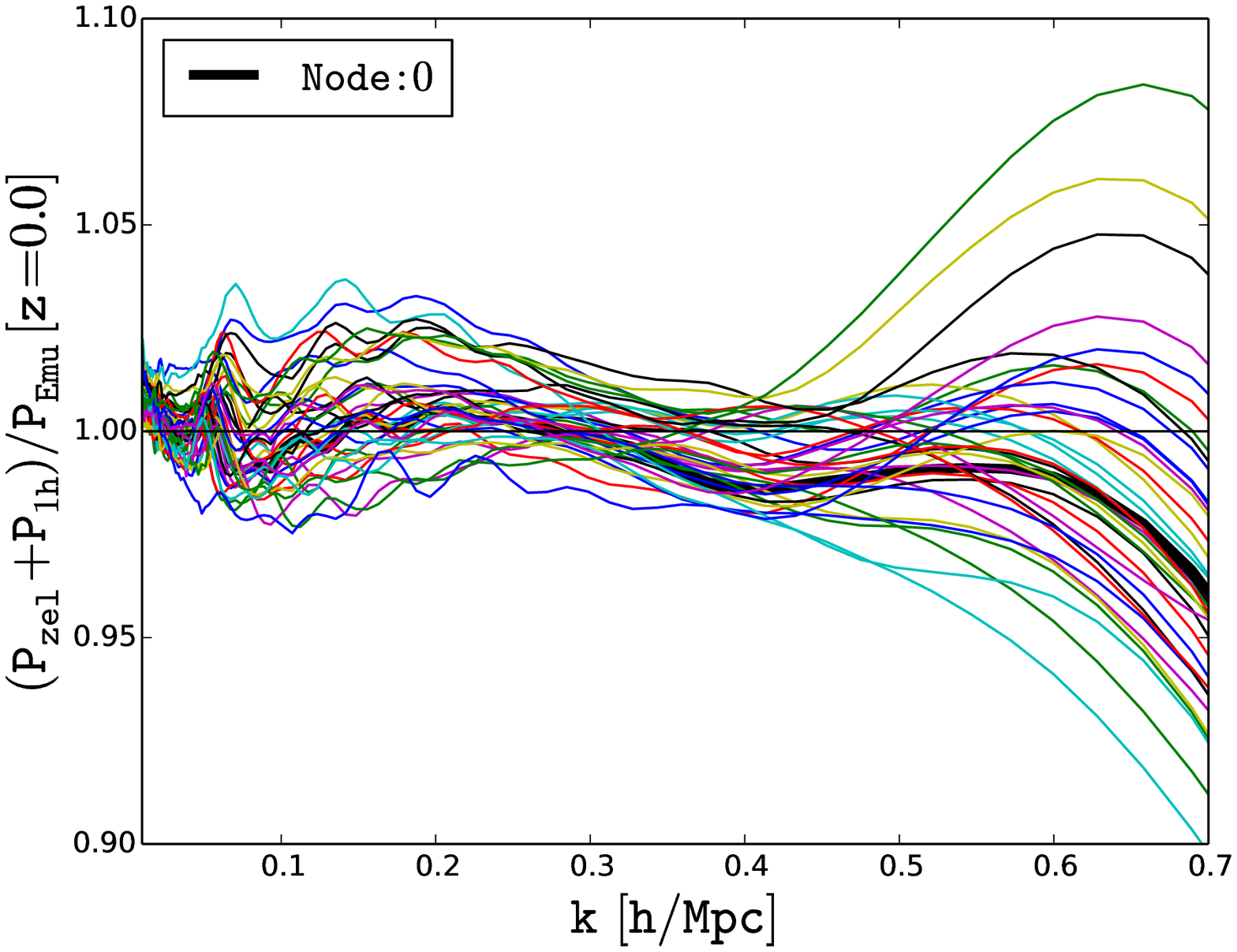}}
\subfigure{\includegraphics[width=0.48\textwidth]{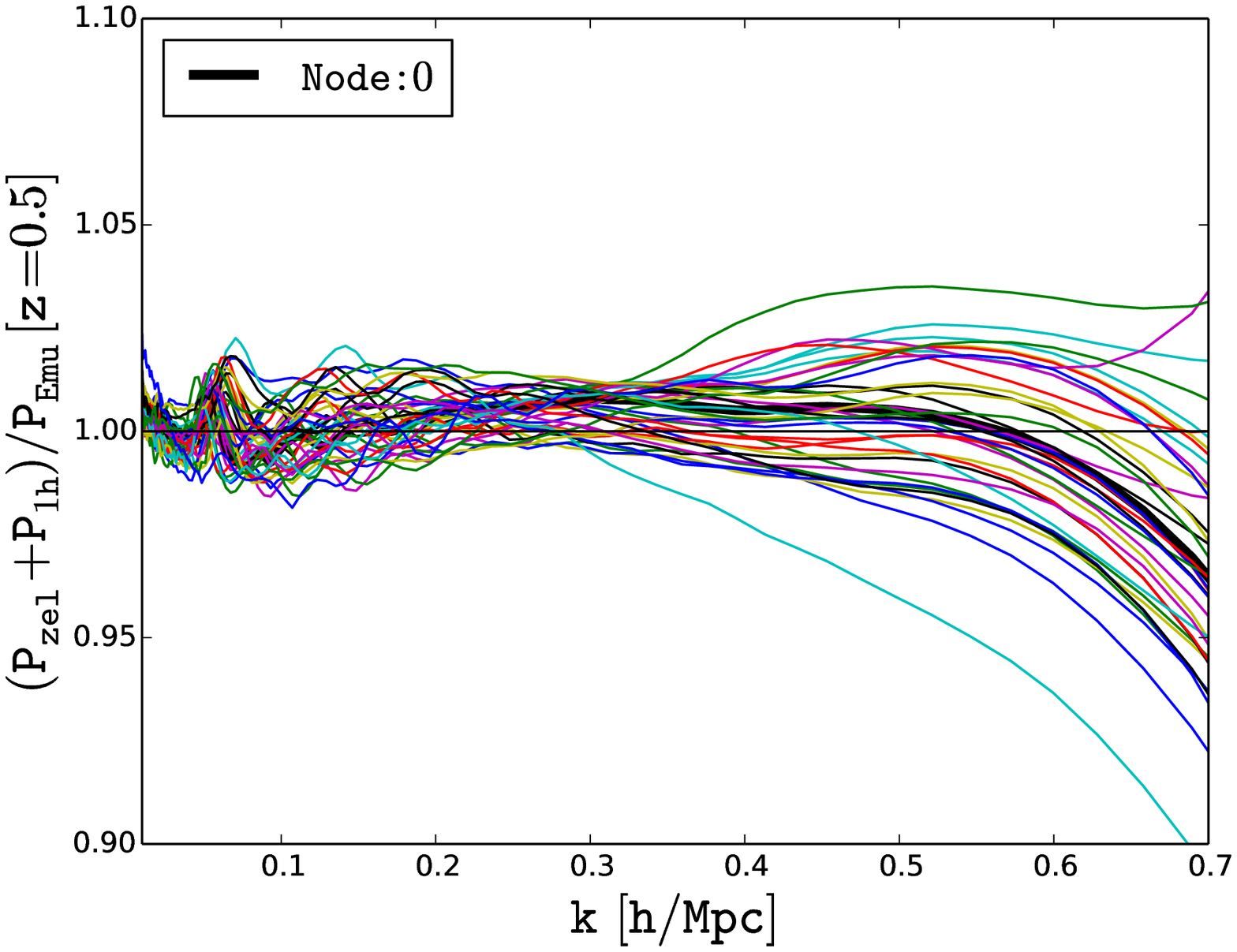}}
\subfigure{\includegraphics[width=0.48\textwidth]{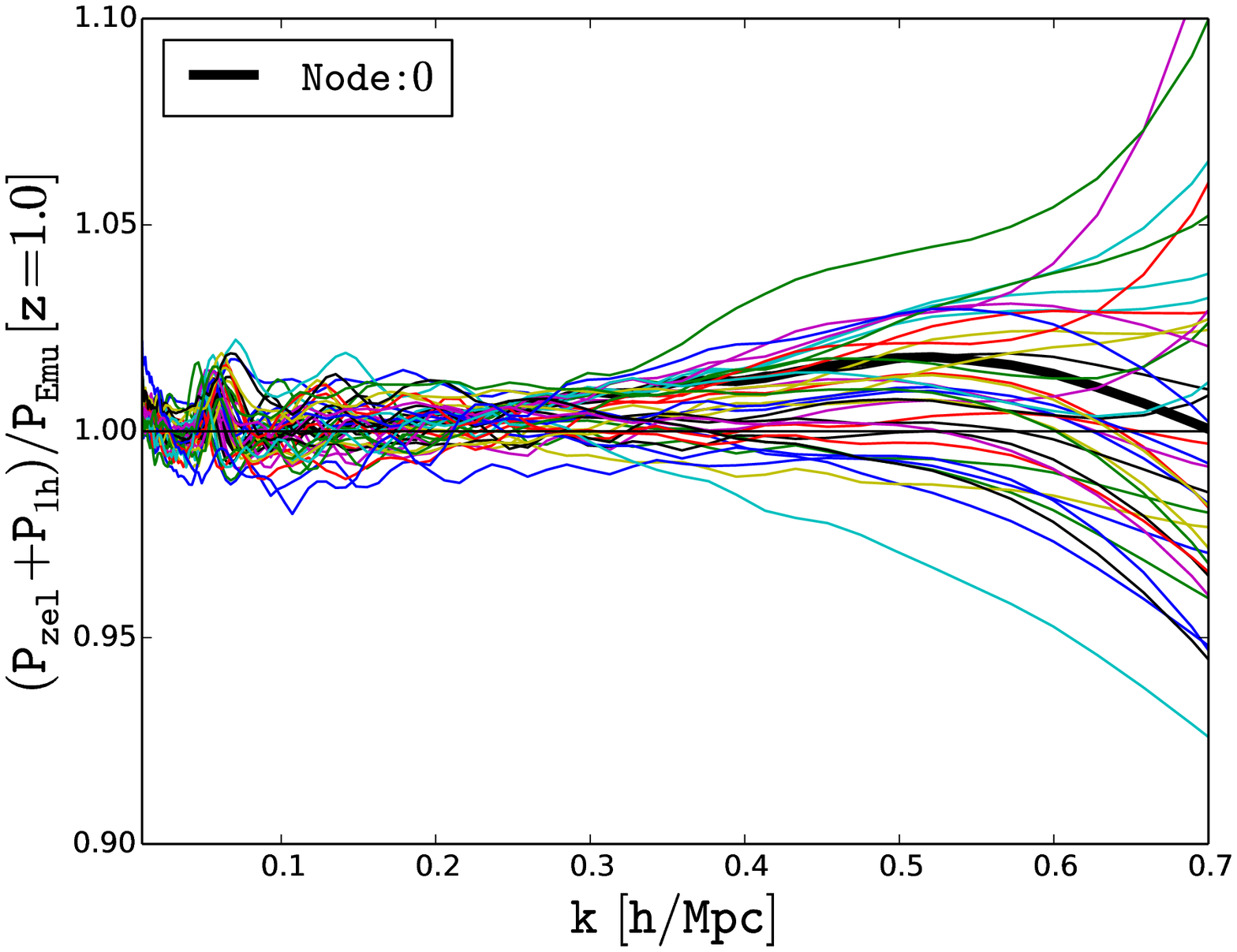}}
\caption{The residuals of our $P(k) = P_{\rm zel}(k)+P_{\rm 1h}(k)$ expression against simulations for 38 different cosmological models (different colour curves in each panel) for three different redshifts.}
\label{fig:final}
\end{figure*}

We tested this expression against the matter power spectrum from emulator ($\rm{P_{emu}}$) on 38 emulator nodes 
where the stated accuracy is 1\%. Figure \ref{fig:final} shows the deviation of our predictions from the true  matter power spectrum of Emulator at three different redshifts: 0.0, 0.5 and 1.0. 

At redshift 0, we can predict the power spectra to a precision of 2$\%$-3$\%$ up to $k \sim 0.5\ h \rm{Mpc}^{-1}$, 
except in some cosmologies which turn out to be unusual (typically equation of state very different 
from $w=-1$). At higher redshifts, this accuracy is even better for the same $k$, as expected since the 
nonlinear effects are smaller. For most of the cosmological models we can calculate these spectra to
5$\%$ up to $k\sim 0.7\ h \rm{Mpc}^{-1}$ and much better for lower $k$.

\begin{figure*}
\centering
\subfigure{\includegraphics[width=0.95\textwidth]{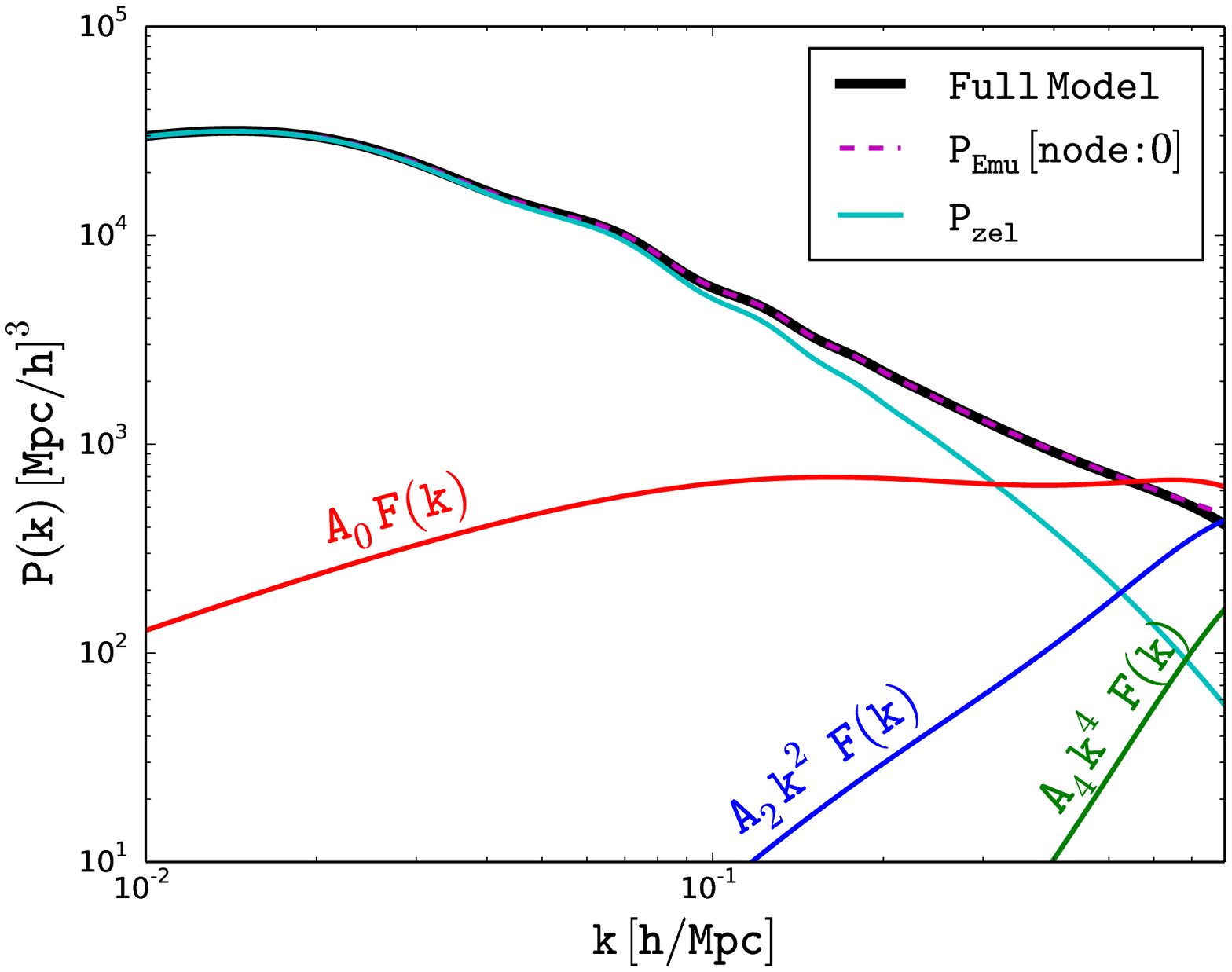}}
\caption{Matter power spectrum for WMAP-7 cosmology at redshift 0.0 from simulations 
(dashed magenta line) versus Zeldovich term (cyan line), $A_0F(k)$ term (red line), $A_2 k^2 F(k)$ term (blue line), $A_4k^4F(k)$ term (green line). 
Thick Black line is the full predicted model from this work and is nearly indistinguishable from the simulations.
}
\label{fig:node0}
\end{figure*}

In figure \ref{fig:node0} we show the prediction of our model for WMAP-7 cosmology (node 0)
with all its components plotted separately. Note that the $A_2k^2F(k)$ (in blue) term has a negative contribution while all other components have a positive contribution. The prediction of node 0 power spectrum is correct to about 2$\%$ up to $k \sim 0.6\ h \rm{Mpc}^{-1}$ increasing to 4$\%$ at $k \sim 0.7\ h \rm{Mpc}^{-1}$. This
can also be seen in figure \ref{fig:final} where thick black line shows the ratio of the predicted and true matter power spectrum for node 0.

\begin{figure*}
\centering
\subfigure{\includegraphics[width=0.48\textwidth]{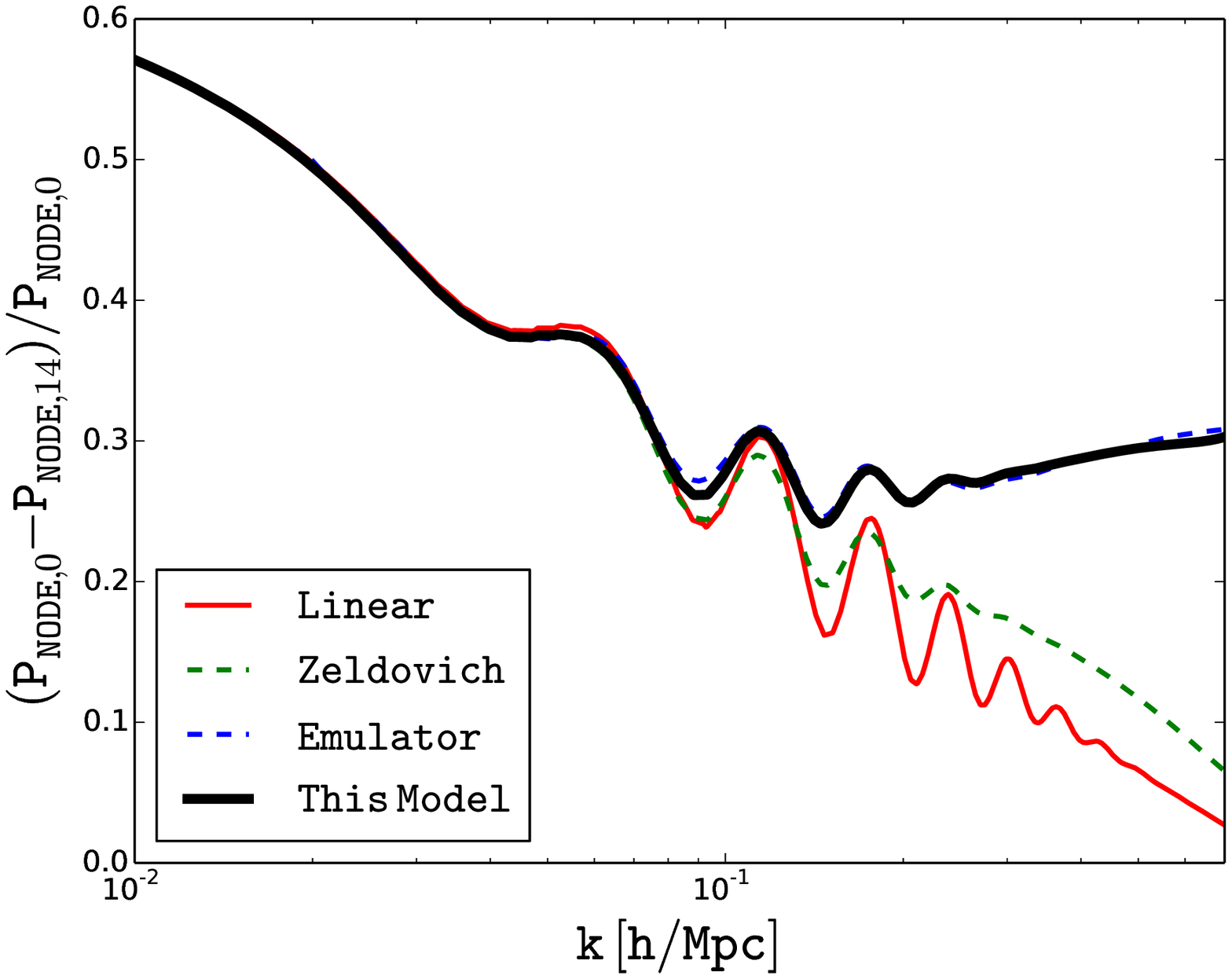}}
\subfigure{\includegraphics[width=0.48\textwidth]{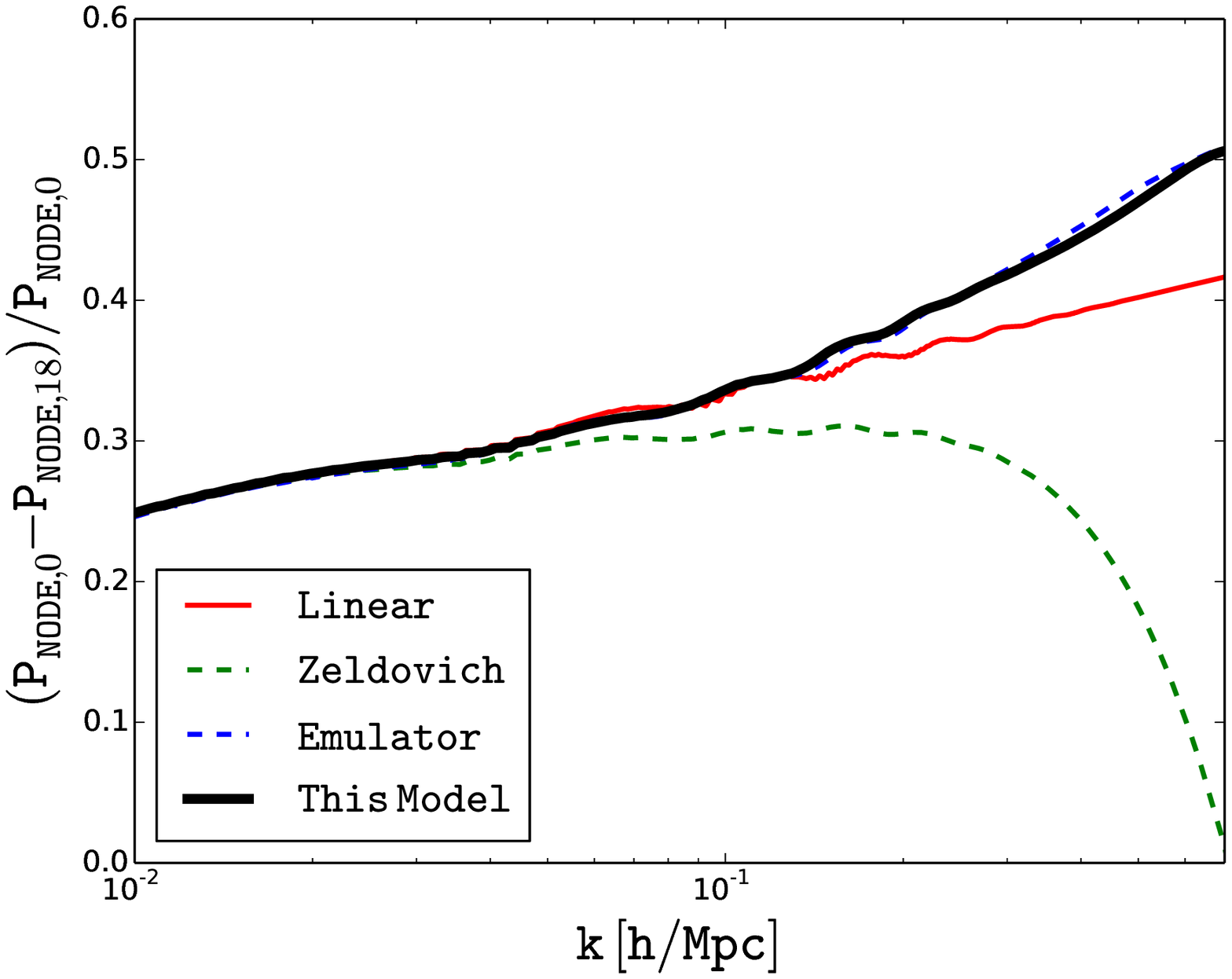}}
\subfigure{\includegraphics[width=0.48\textwidth]{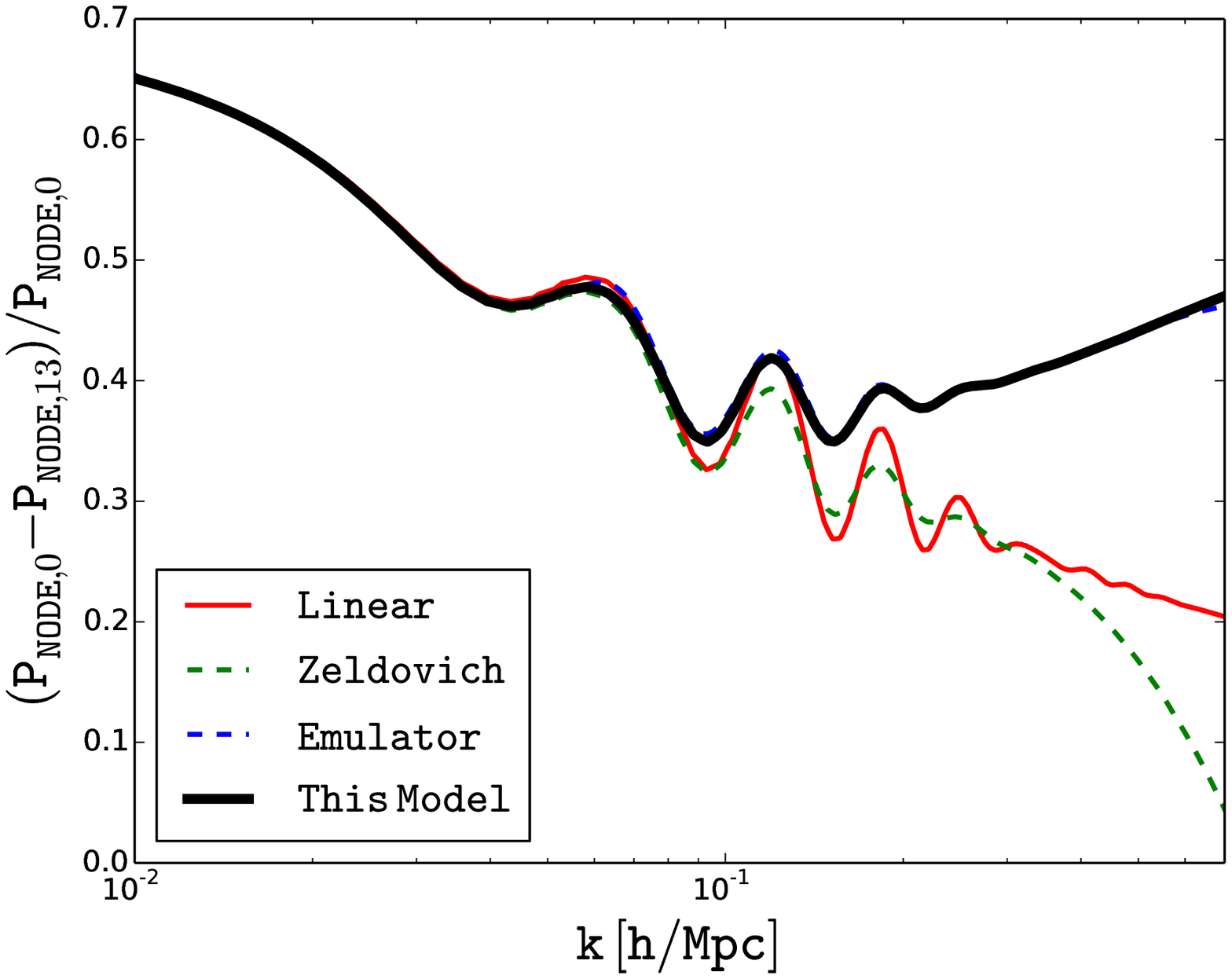}}
\subfigure{\includegraphics[width=0.48\textwidth]{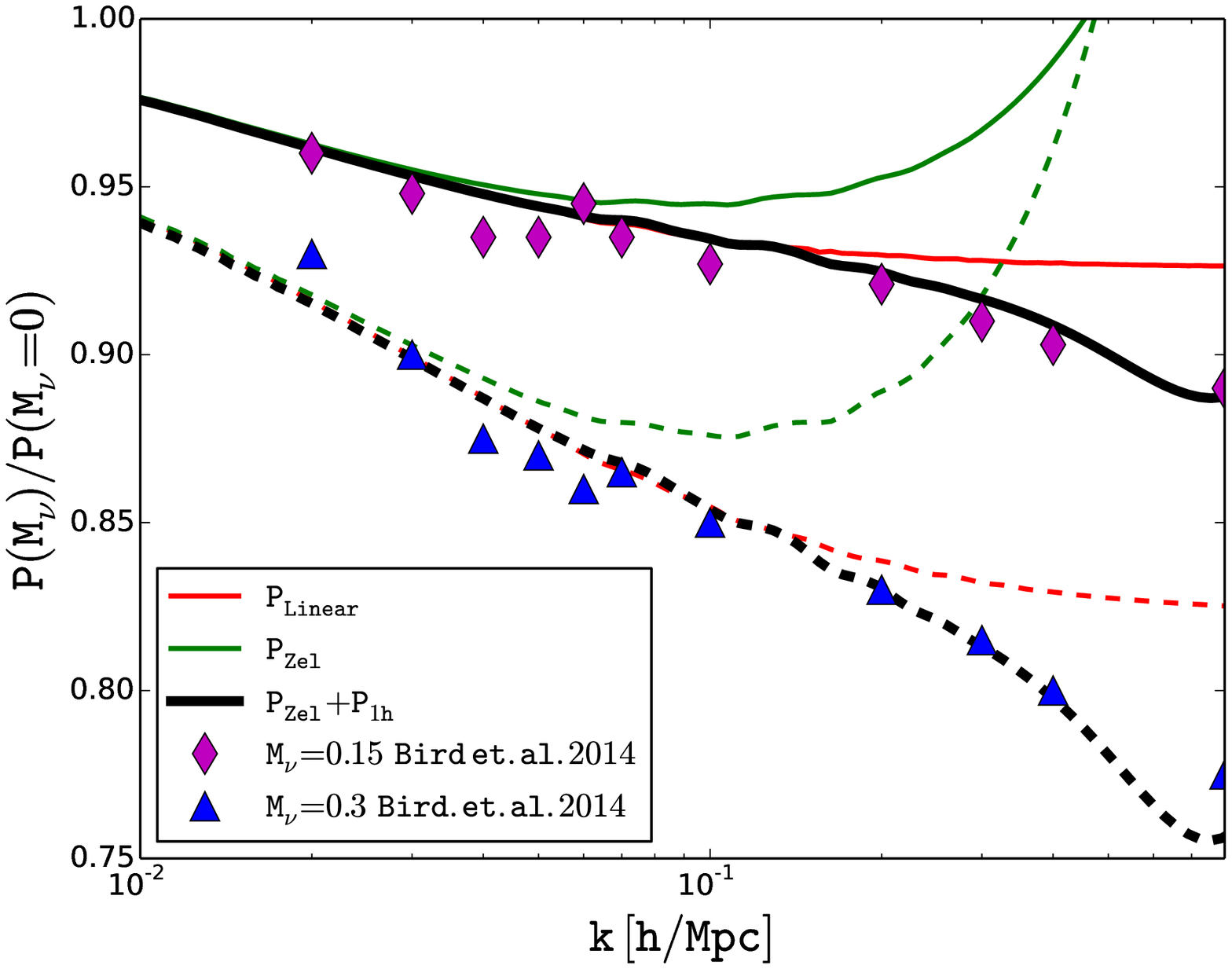}}
\caption{Relative difference in matter power spectrum between node 0 (Emulator) and node 14 (top-left), 18 (top-right), 13 (bottom-left). Showing the same quantity for linear term (in solid red), Zeldovich term (in dashed green), Emulator power spectrum (in dashed blue) and our prediction (in solid black). Bottom-right panel shows the ratio of the matter power spectrum with and without neutrino mass, for $\sum M_{\nu}$=0.15 (in solid lines) and 0.3 (in dashed lines) from Bird et. al. (2012).}

\label{fig:derivatives1}
\end{figure*}

We also explored how well can this expression predict the changes in the matter power spectrum when 
cosmological parameters are changed. We take emulator node 0 as the fiducial model and plot the relative difference with other nodes. The first three panel of figure \ref{fig:derivatives1} (in reading order) shows these derivatives for different components: linear term (in red), Zeldovich term (in green), emulator (in blue) and our predicted model (in thick black). Our predictions are matching very well with that of the true matter power spectrum from emulator, and certainly much better than pure linear theory or pure Zeldovich approximation. Note that we 
also get very good agreement of BAO smoothing, in contrast to linear theory predictions: this is because we 
are using Zeldovich approximation which smears out BAO. 
The broadband effects of Zeldovich approximation are often anti-correlated with $A_0$: this is because an 
increase in $\sigma_8$ increases the nonlinear smearing caused by the linear streaming of the 
displacement field, reducing the 
amplitude of the power spectrum in the 
Zeldovich approximation, while at the same time the amplitude of the $A_0$ is increased by the 1-halo 
term, generated by having more halos at the same halo mass. 
The latter effect typically wins: the total power spectrum and the Zeldovich power spectrum are typically, 
but not always, on the opposite side relative to the linear power spectrum. 

Of particular interest is the change in neutrino mass, also 
shown in figure \ref{fig:derivatives1}. We compare the model 
predictions to the simulations of \cite{2012MNRAS.420.2551B}. We see that our model predicts nearly perfectly the changes in
the nonlinear power spectrum induced by massive neutrinos. This shows that nonlinear effects of 
massive neutrinos are no different than any other parameter: on large scales they follow linear theory, 
while on small scales the effects are dominated by the change in $A_0$. For $\sum m_{\nu}=0.15eV$ 
the change in $\sigma_8$ is about 3\% and the corresponding change in $A_0 \propto \sigma_8^{3.9}$ 
is 13\%, while Zeldovich approximation goes in the opposite direction, so the 
linear suppression of 7\% at $k \sim 0.2\ h \rm{Mpc}^{-1}$ is increased to 11\% at $k \sim 0.8\ h \rm{Mpc}^{-1}$, 
in a perfect agreement with simulations.

\section{Covariance matrix and the cosmological information content of $P(k)$}
\label{sec:covariancematrix}

We next turn to the issue of covariance matrix. On large scales, low $k$, the covariance matrix is based on
Gaussian approximation. 
As we move to higher $k$ the modes become correlated and the covariance matrix becomes non-Gaussian. 
In our model the non-Gaussianity comes from two separate terms. First is the non-Gaussian nature of the 
Zeldovich term and second is the non-Gaussian nature of the 1-halo term. 
We will not analyze the non-Gaussian covariance matrix in Zeldovich approximation in this paper, 
as there are currently no analytic calculations available. 
We also do not have any analytic predictions for the correlation 
between the Zeldovich part and the 1-halo part. For the 1-halo term we will focus on $A_0$ contribution, since as we will argue in next section we should marginalize over the higher order terms anyways. In our 
initial discussion we will
ignore the super-sample variance contribution \citep{2013PhRvD..87l3504T,2014PhRvD..89h3519L}, 
which will be discussed separately below. 

The halo model calculations in figure \ref{fig:variance} suggest that the relative variance $\sigma_{A_0}/A_0$
should be around 0.01$\sqrt{({\rm Gpc}/h)^3/{\rm Volume}}$, depending on the cosmological model and 
redshift. 
This calculation is given by 
\begin{equation}
\left({\sigma_{A_0} \over A_0}\right)^2={\int f(\nu)d\nu M^3 \over [\int f(\nu)d\nu M]^2\bar{\rho} V},
\end{equation}
and is determined by the 4th moment of mass integrated over
the halo mass function and thus very sensitive to the halo mass function 
accuracy at the high mass end. 
Just as in the case of the halo model predictions for the scalings of $A_0$, $A_2$ and 
$A_4$, we may not completely trust the halo model predictions. 
We will write the following ansatz to the 
covariance matrix ${\rm Cov}(P(k_i),P(k_j))=\langle P(k_i)P(k_j)-\langle P(k_i) \rangle \langle P(k_j)\rangle$, 
\begin{equation}
{\rm Cov}(P(k_i),P(k_j))=P(k_i)P(k_j)\left({2 \over N_i}\delta_{ij}+\left({\sigma_{A_0} \over A_0}\right)^2\right).
\label{cov}
\end{equation}
Here $N_i$ is the number of Fourier modes in the $i-$th bin. Our model predicts that the scaling of the variance is
\begin{equation}
{\sigma_{A_0} \over A_0}={\delta_{A_0} \over [(V/1 h^{-1} {\rm Gpc})^3]^{1/2}}, \,\, \delta_{A_0}=0.0079({h^{-1} {\rm Gpc})^{3/2}},
\label{sigaa}
\end{equation}
where $V$ is the volume in units of $(h^{-1} {\rm Gpc})^3$, and the value of $\delta_{A_0}=0.0079({h^{-1} {\rm Gpc})^{3/2}}$
was obtained from a
fit of the model to the diagonal part of the covariance matrix derived from Planck cosmology
simulations in \cite{2014PhRvD..89h3519L}, shown in the right panel of figure \ref{fig:ssv}. 
This value is slightly lower than 
the predictions of the halo model in figure \ref{fig:variance}. Since the predictions are 
very sensitive to the massive end of the halo mass function, which is not well determined, we should not expect a perfect
agreement. 

It is important to note that the covariance matrix depends on the simulated volume: if the volume changes the 
covariance matrix will change, and this means that comparing one set of covariance matrix results to another is 
not trivial. We can simplify the expression if we express the number of modes in terms of a fixed width of the 
$k$ bin $\Delta k$, $N=4\pi k^2\Delta k V/(2\pi)^3$. One can see that both
the gaussian sampling variance term and the Poisson term scale with volume, so that 
\begin{equation}
{\rm Cov}(P(k_i),P(k_j))=P(k_i)P(k_j)V^{-1}\left({4\pi^2 \over k_i^2\Delta k}\delta_{ij}+\delta_{A_0}^2\right).
\label{cov1}
\end{equation}
The relative contribution of diagonal versus off-diagonal terms still depends on the width of the binning in $k$, 
but the overall volume scaling is the same. 

\begin{figure*}
\centering
\subfigure{\includegraphics[width=0.48\textwidth]{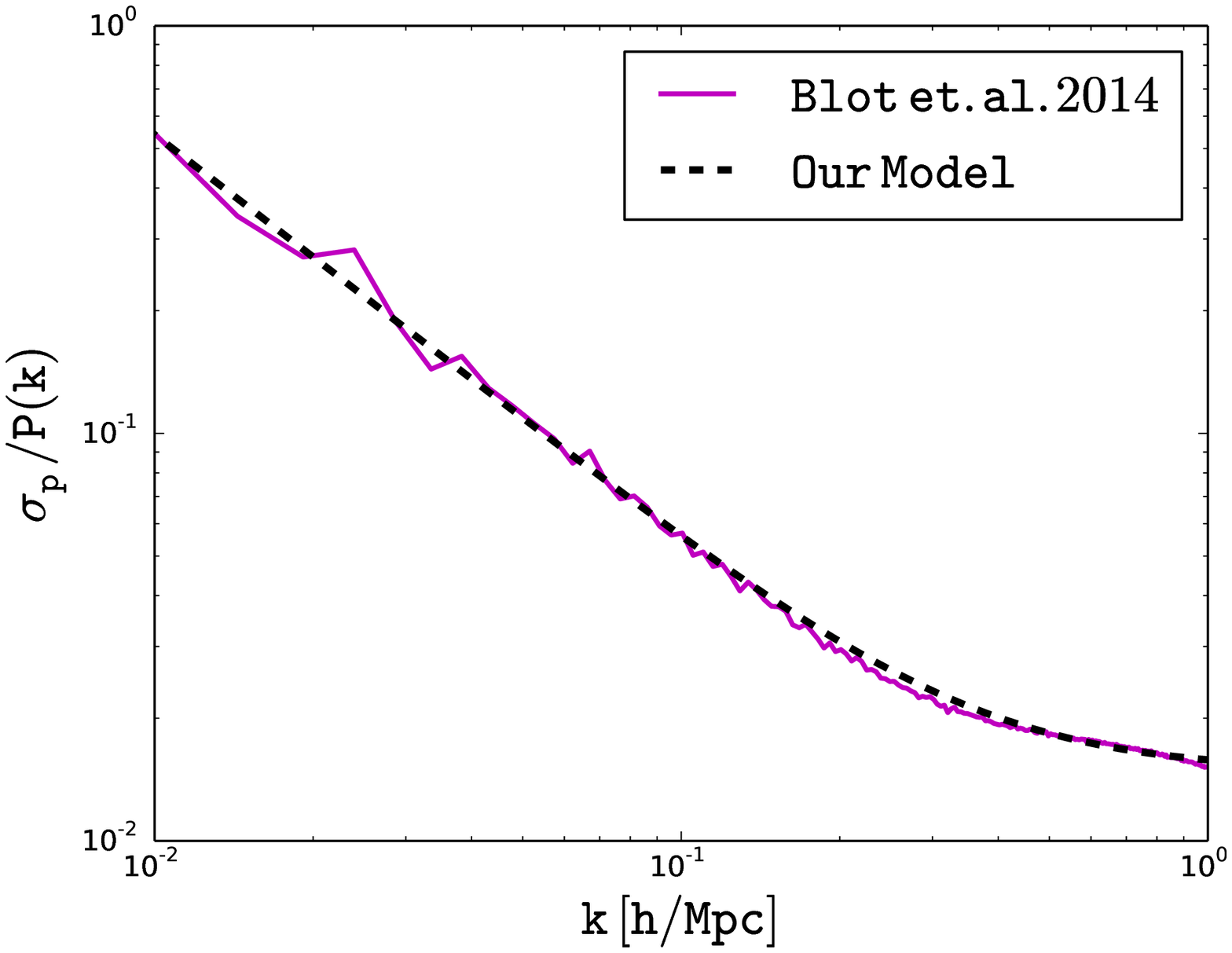}}
\subfigure{\includegraphics[width=0.48\textwidth]{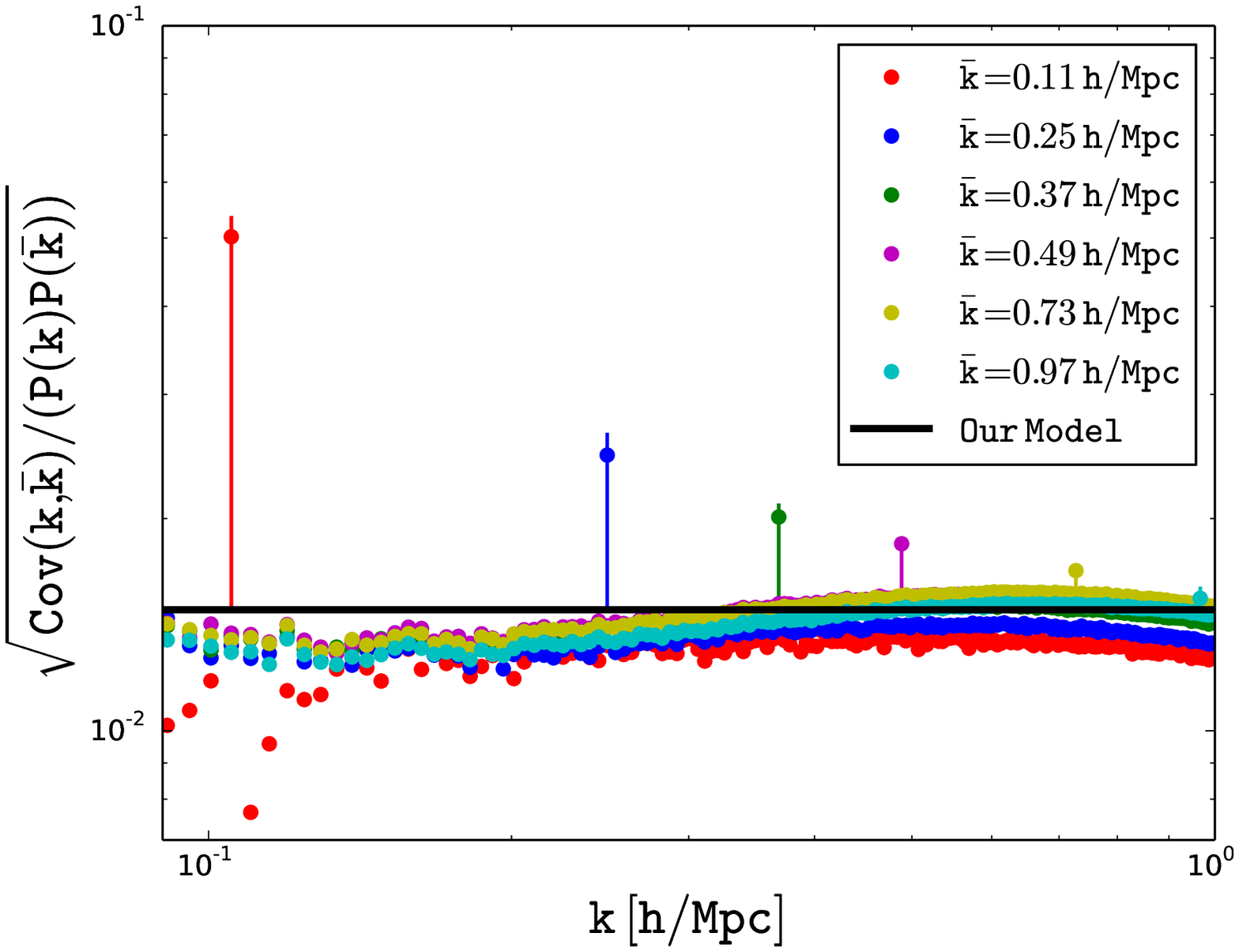}}
\subfigure{\includegraphics[width=0.48\textwidth]{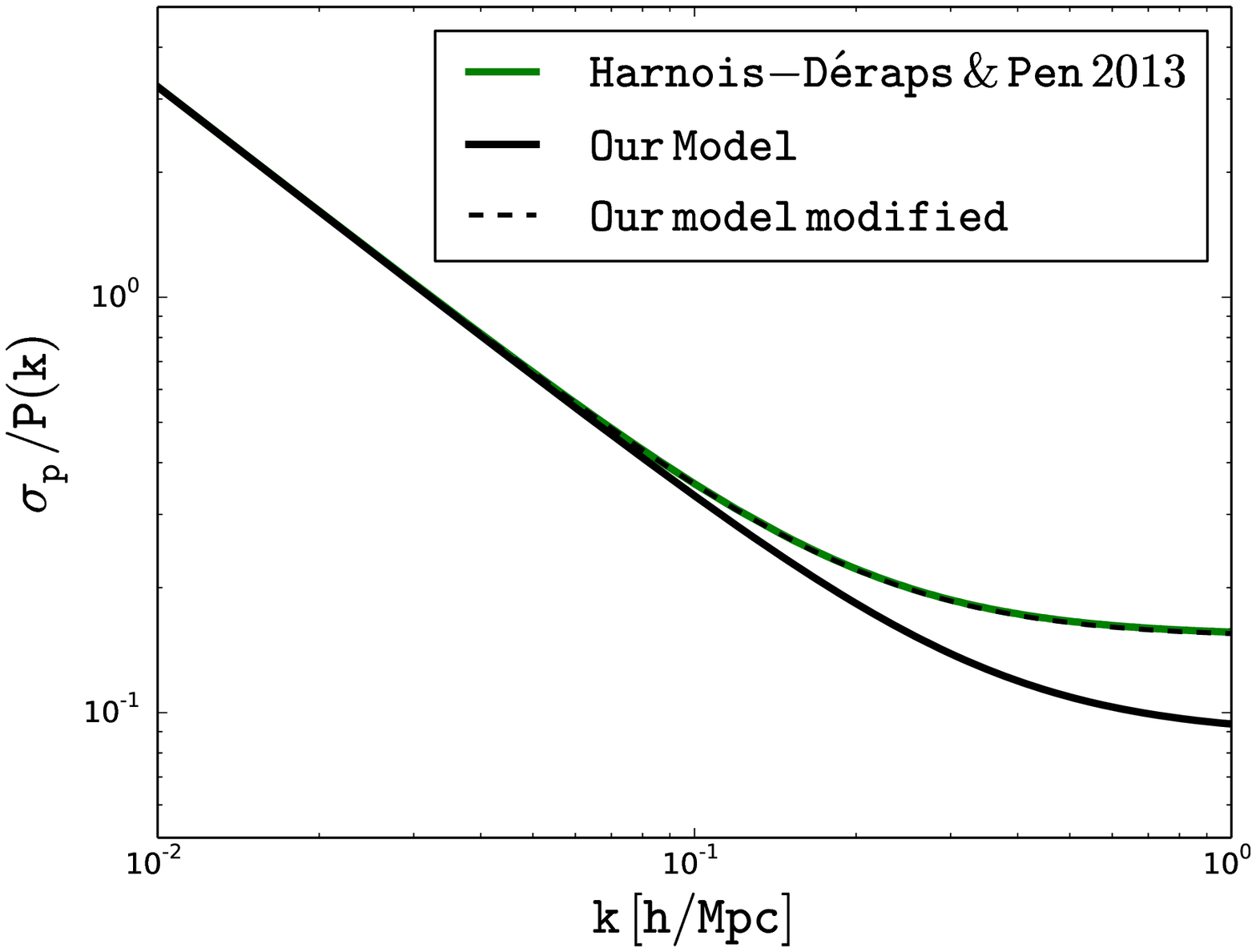}}
\subfigure{\includegraphics[width=0.48\textwidth]{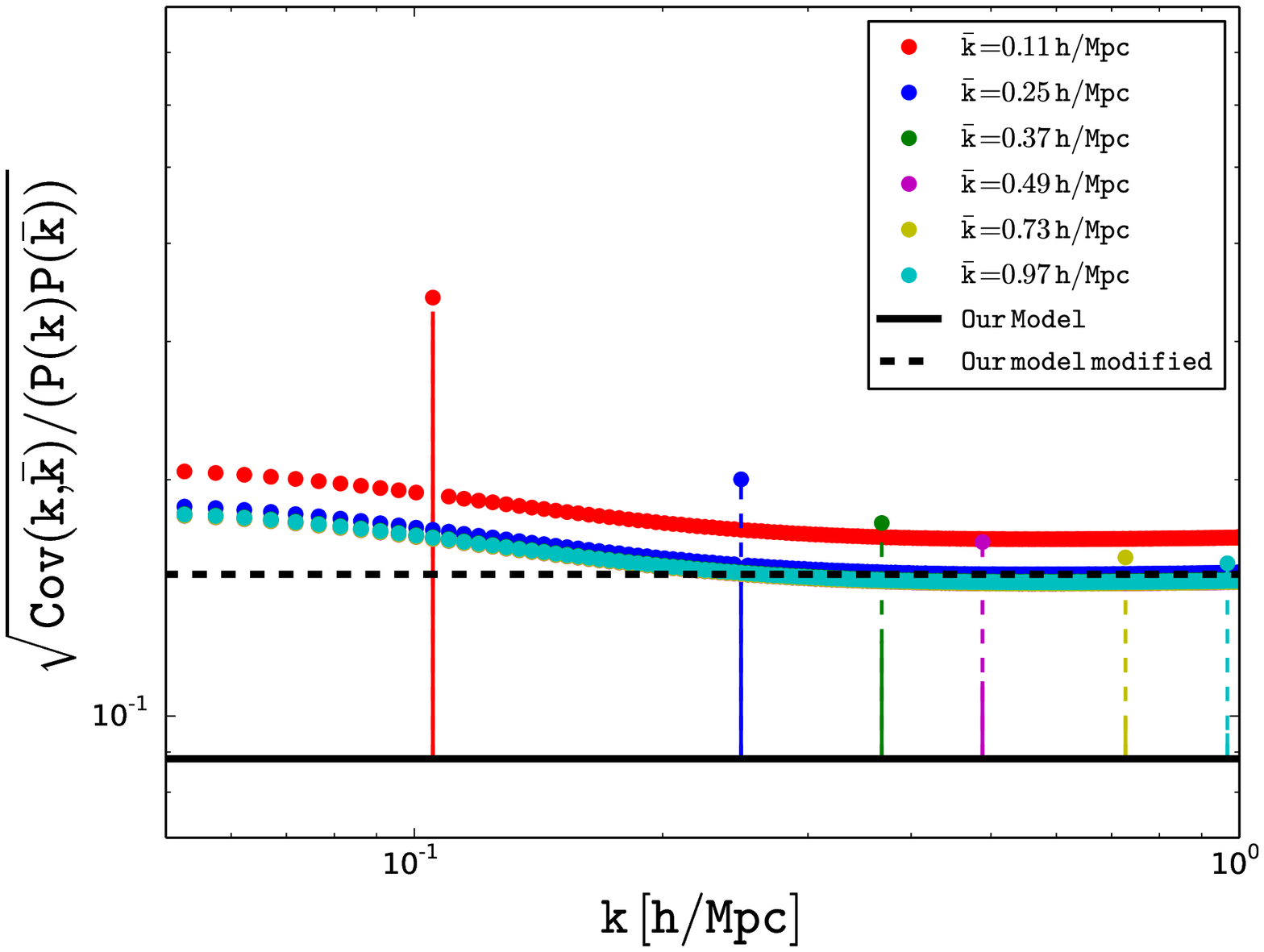}}
\caption{Comparison between our model prediction of covariance matrix with Blot et. al. 2014 (upper panels) and Harnois-D{\'e}raps \& Pen 2013 (lower panels) for diagonal (left panels) and off-diagonal elements (right panels). Note that there are no free parameters in the top, while for the bottom panel we show both our best model without a free parameter as well as a modified 
model where we fit for the value of $\sigma_{A_0}/A_0$, which is a valid procedure for these simulations, 
as discussed in the text. 
Our covarince matrix model (equation \ref{cov}) is very simple, yet it is able to reproduce the full covariance matrix from 
simulations to within 10-20\%.} 
\label{fig:linda}
\end{figure*}

Now that we have fixed the only free parameter of our model $\delta_{A_0}$ 
we can apply it to another set of simulations to see 
the agreement. We have compared it to results in \cite{2014arXiv1406.2713B}, which 
used 12288 boxes of size 656.25$h^{-1} \rm{Mpc}$ to derive the full covariance matrix.  
In figure \ref{fig:linda} (upper panels) we have compared our model to these simulations 
for both diagonal and off-diagonal parts of the covariance matrix.
We show that 
the diagonal part of the covariance matrix (left panel of figure \ref{fig:linda}) is an excellent fit, even 
better than comparison with \cite{2014PhRvD..89h3519L}, and this is without any free parameters. 
In the right panel of figure \ref{fig:linda} we show the off diagonal terms for six different $k$ values. 
Our model predicts the off-diagonal correlation coefficients are simply a constant, except at the diagonal where 
there is an additional Gaussian contribution. 
Our prediction is in a reasonable agreement with these simulations: 
we are able to reproduce simulation results for both diagonal and off-diagonal terms to within 10-20\%, which is 
remarkable given its simple form and no free parameters. 
 
\subsection{Variance of the covariance matrix}

\begin{figure*}
\centering
\subfigure{\includegraphics[width=0.8\textwidth]{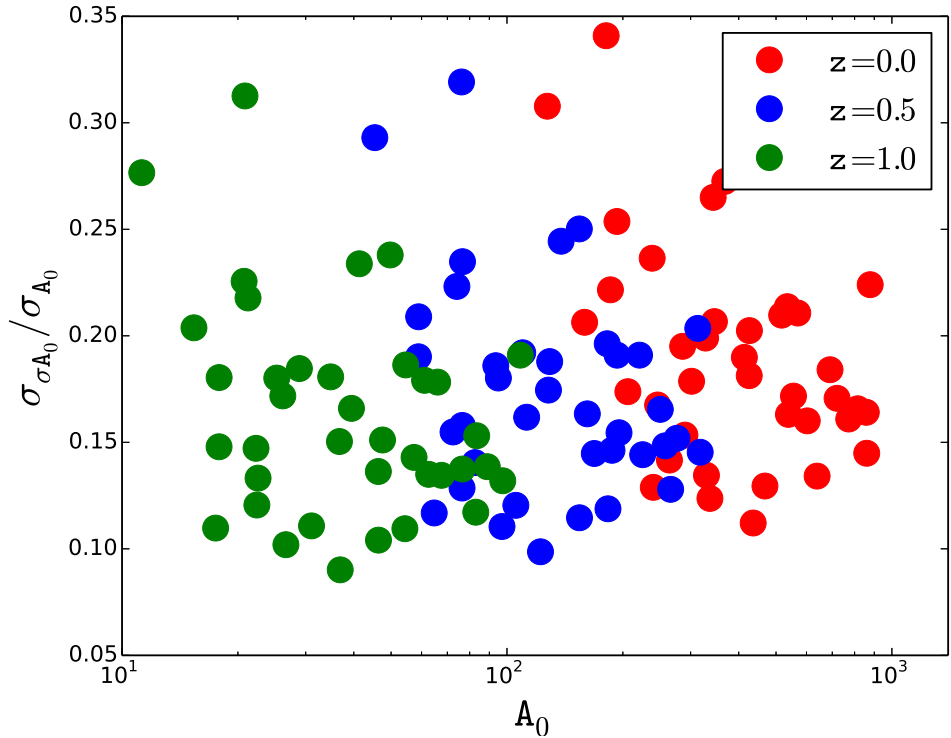}}
\caption{Relative variance $\sigma(\sigma_{A_{0}})/\sigma_{A_{0}}$ versus $A_0$ based on our model 
for three different redshifts: 0.0 (red), 0.5 (blue) and 1.0 (green). Each bullet is one cosmological realization of the 38 cosmic emulator nodes. 
 }
\label{fig:varvarA0}
\end{figure*}

An interesting and important question is how big do the simulations need to be to converge. For the convergence of 
the power spectrum the answer is given by $\sigma_{A_0}/A_0=\delta_{A_0}/V^{1/2}$ and we can see that $V=1(h^{-1}{\rm Gpc})^3$
is sufficient for 1\% accuracy. For the covariance matrix this requirement becomes considerably stricter. One 
can write an expression for the relative variance of the covariance term as 
\begin{equation}
\left({\sigma(\sigma_{A_0}) \over \sigma_{A_0}}\right)^2={\int f(\nu)d\nu M^7 \over [(\int f(\nu)d\nu M^3]^2\bar{\rho} V},
\end{equation}
so we can see that this is given by the 8th moment of the mass averaged over the halo mass function. 
The results of this prediction are shown in figure \ref{fig:varvarA0}. 
The rms variance for $V=1(h^{-1}{\rm Gpc})^3$ is now about 10-30\% and the corresponding error on the 
covarince matrix (which goes as a square of $\sigma_{A_0}$) is thus 20-70\%. There is a large spread in the value because the 
calculation is so sensitive to the very high mass end of the halo mass function, which is poorly known, so the 
resulting values should only be taken as indicative and can probably vary by a factor of 2. This is simple to 
understand: occasionally there will be a large cluster formed which will significantly change the value of $A_0$, 
and consequently make its variance change considerably. 

As an example, when we compare our model predictions of the covariance matrix to \cite{2012MNRAS.423.2288H} we find that the 
agreement is not very good, in that our model predicts lower covariance matrix than measured, and the predicted value 
of $\sigma_{A_0}/A_0$ is about 40\% below the required for fit the simulations. 
However, \cite{2012MNRAS.423.2288H} used a total simulated volume of $1.6(h^{-1}{\rm Gpc})^3$, suggesting that the 
value of $\sigma_{A_0}/A_0$ has only been determined to about 10-25\%. If we let the value of $\sigma_{A_0}/A_0$
to be free, we again find a remarkable agreement with the simulations. 

To converge on the covariance matrix at 1\% one needs a simulated volume to be of order $500-5000(h^{-1}{\rm Gpc})^3$. 
This is an enormous volume: it explains why in recent work of \cite{2014arXiv1406.2713B} they needed to simulate 
12288 simulations with a total volume of $3350(h^{-1}{\rm Gpc})^3$ to converge. 

\subsection{Information content}

We can now combine the variance of $A_0$ with its scaling with $\sigma_8$, $A_0 \propto \sigma_8^{3.9}$, 
to derive the cosmology information 
content of the $A_0$ term,
\begin{equation}
\dfrac{\sigma_{\sigma_8}}{\sigma_8}=\dfrac{\sigma_{A_0}}{3.9 A_0}=0.002 \sqrt{(h^{-1}{\rm Gpc})^3/{\rm Volume}}.
\label{eqn:var_s8}
\end{equation}
This is a remarkably small number, which suggests that much of the cosmological information on the 
rate of growth of structure, and consequently on the Figure of Merit for dark energy equation of state 
\citep{2010PhRvD..82f3004M}, resides in this term. To achieve a comparable precision on linear scales one would 
need about $5\times 10^5$ modes, which for $1(h^{-1}{\rm Gpc})^3$ volume 
would correspond to $k_{\rm max}=0.31\ h \rm{Mpc}^{-1}$. 
This is already well into the nonlinear regime for $z<1$ implying that we do not have this number of linear modes available, 
so the bulk of the cosmological information on the amplitude comes from $A_0$ term. 
However, since $A_0$ is mostly sensitive to amplitude (best correlation is with $\sigma_{11.3}$) 
and nothing else, this also suggests that information 
on other parameters that depend on the shape of $P(k)$ and not its amplitude will be less well 
determined. 

While we do not have reliable variance predictions for $A_2$ and $A_4$ from simulations, 
figure \ref{fig:variance} suggests that $A_2$ 
has variance 3 times larger than $A_0$ and $A_4$ has variance another 3 times larger than $A_2$. 
This is mostly caused by the fact that Poisson fluctuations get larger for higher order coefficients because 
of their mass weighting: for example, $A_2$ weighting is $M^{5/3}$ as opposed to $M$ for $A_0$, giving more 
weight to higher mass halos, which are rarer and therefore have larger Poisson fluctuations. This, 
combined with less steep scaling of $A_2$ and $A_4$ with $\sigma_8$ compared to $A_0$ (equation \ref{eqn:a0}, \ref{eqn:a2} and \ref{eqn:a4}), 
suggests that there is little additional information in these two coefficients. 
Another argument for why information in $A_2$ and $A_4$
should be ignored, based on baryonic effects, will be presented below.

\begin{figure*}
\centering
\subfigure{\includegraphics[width=0.48\textwidth]{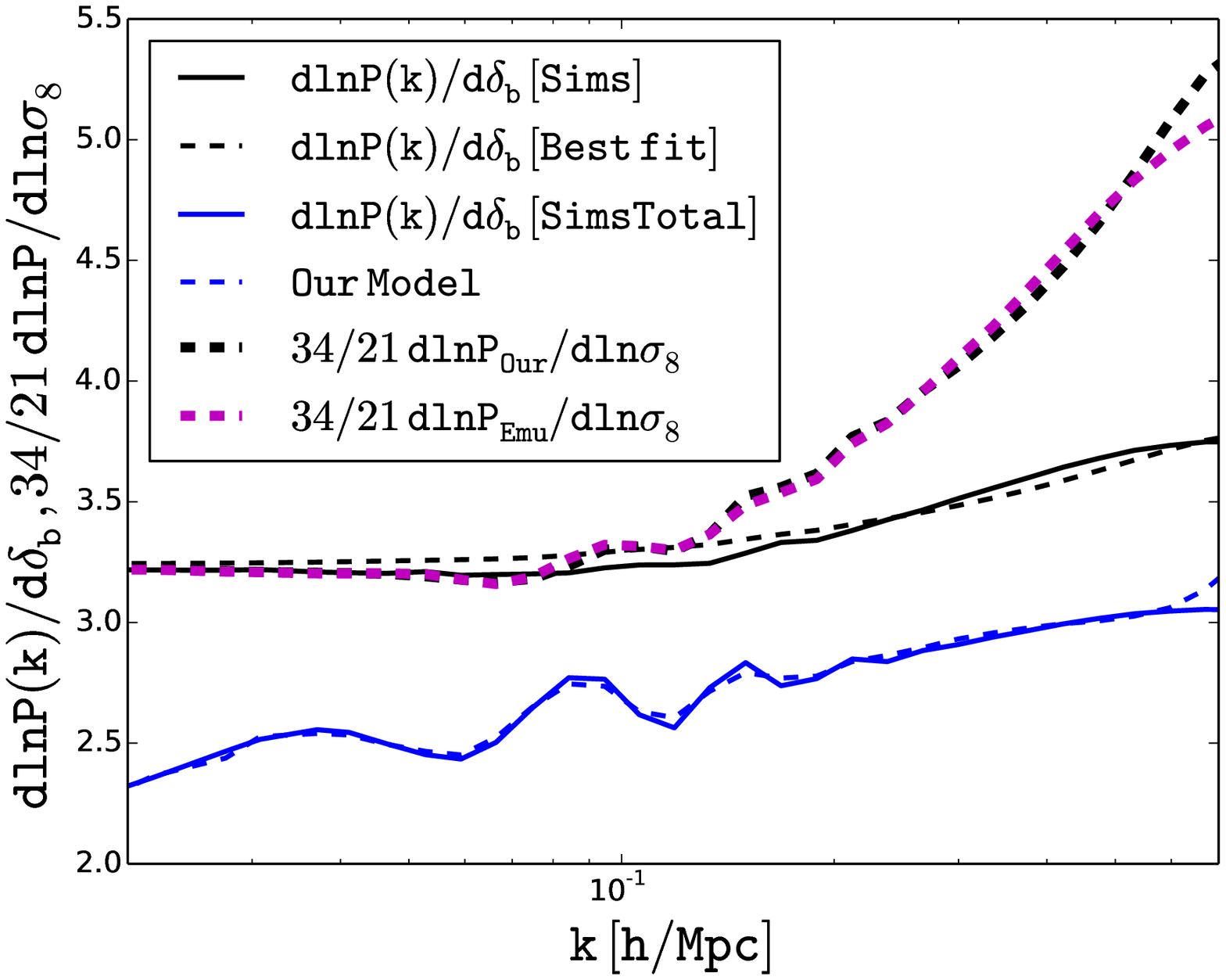}}
\subfigure{\includegraphics[width=0.48\textwidth]{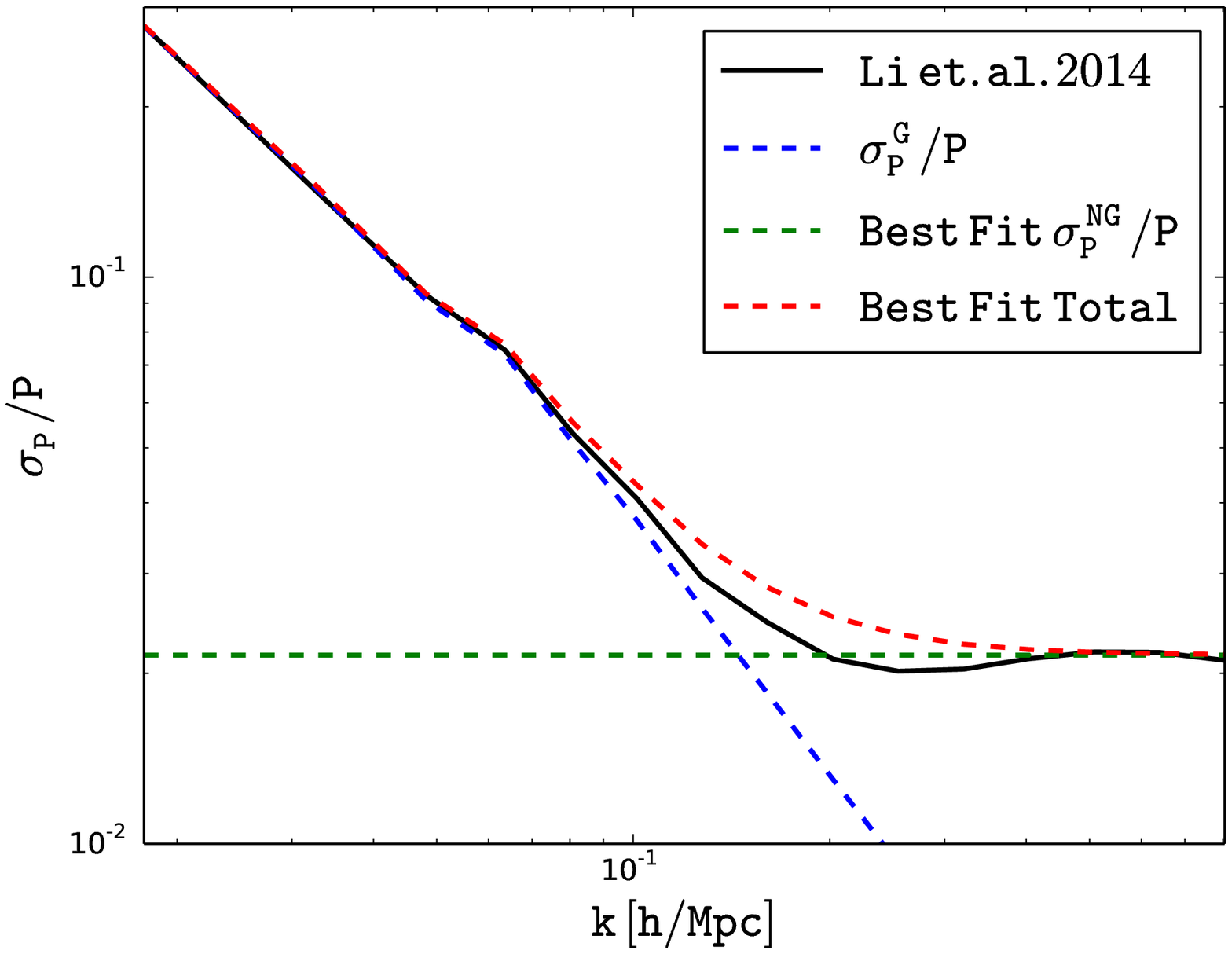}}
\caption{Left: Derivative of the matter power spectrum with respect to the 
change in curvature (i.e. background density) 
from simulations (blue solid line) and our best
fit model (blue dashed line). The same, but only for the growth effect without dilation, 
is shown with the corresponding black solid and 
black dashed lines. 
The thick dashed lines show the derivative with respect to the amplitude change, such that it is degenerate 
with the curvature change at low $k$: shown are the predictions from simulations
(in magenta) and from our model (in black). We see that the degeneracy is broken at higher $k$ even in the 
absence of the dilation effect. 
Right: Relative variance in the matter power spectrum:  $\sqrt{2/N}$ (blue-dashed line) where $N$ is the number of modes, best fit $\sigma_p/P$ (green-dashed line), and the total (red-dashed line) as the norm of the two terms. }
\label{fig:ssv}
\end{figure*}

\subsection{Super-sample variance}\label{sec:ssv}

Super-sample variance \citep{2006MNRAS.371.1188H} arises from the 
very long wavelength density modes that appear as constant on the scale of the survey. 
These can be viewed as a change of curvature inside the 
observed volume \citep{2011JCAP...10..031B}, and this couples to all the short wavelength modes. On large scales 
the effect can mimic a change in the amplitude of fluctuations, together with a rescaling of the 
length \citep{2012PhRvD..85j3523S}: 
\begin{equation}
\delta \ln P(k)=\left({47 \over 21}-{1 \over 3}{d\ln P \over d\ln k}\right)\delta_b = \left({68 \over 21}-{1 \over 3}{d\ln (k^3P) \over d\ln k}\right)\delta_b,
\label{eqn:ssv}
\end{equation}
where $\delta_b$ is the density perturbation on the scale of the survey volume. 
The first term is the effect of the curvature on the growth of small scale modes, 
while the second term is the dilation due to the presence of local curvature. 
It is important to recognise that on large scales the growth 
effect is degenerate with a $(34/21) \delta_b$ change of 
amplitude $\sigma_8$, while the dilation effect of
$-\delta_b/3$ is degenerate with a change in scale, i.e. with a change in 
the angular diameter distance that can arise from a change in cosmological parameters. 
We will assume that the change in scale cannot be used as an indicator of the 
super-sample variance because of its degeneracy with these other parameters, so we will only focus on 
the change in growth rate. 
The rms fluctuations of $1({\rm Gpc}/h)^3$ volume are about 0.4\% \citep{2013PhRvD..87l3504T}, 
which together with the $34/21$ factor implies that at low $k$ one cannot determine $\sigma_8$ 
to better than 0.6\% in the linear regime, which is a factor of 3 larger error than the error on $\sigma_8$ 
without the super-sample variance in equation \ref{eqn:var_s8}. It is therefore clear that without addressing this issue the 
super-sample variance dominates the errors. 

On smaller scales we expect the nonlinear effects  are no longer degenerate with a change in $\sigma_8$. 
Physically the reason for difference is in the curvature nature of the super-sample
variance: curvature effects grow with the growth rate, i.e. the growth of short wavelength mode $\delta_s$ due to the 
coupling to the long wavelength mode
scales as $\delta_s(z)[1+34D(z)\delta_{b0}/21]$, where $D(z)$ is the linear growth rate and $\delta_{b0}$ is the long 
wavelength mode today,  and thus this coupling only matters
at low redshifts since $D(z)\ll 1$ for $z\gg 1$. This
is different from a simple change in overall amplitude $\delta_s(z)(1+\delta \sigma_8)$, which has no redshift dependence. 

To understand this more quantitatively we can compute the logarithmic 
derivative of $A_0$ (equation \ref{A024}) with respect to the 
two parameters in the context of the universal halo mass function $f(\nu)$, where $\nu$ is given by equation \ref{nu}
\citep{2008JCAP...08..031S}. 
The Lagrangian bias is defined as $b_L=\bar{n}^{-1}\partial n/\partial \delta_b$, which can be rewritten using 
$\nu=(\delta_c-\delta_b)^2/\sigma^2$ as $b_L=(-2\nu/\delta_c)\partial \ln[\nu f(\nu)]/\partial \nu$. 
In addition we also have the mean density increased by $\delta_b$ inside the patch. We are still dividing
the density with the 
global mean density, so $\bar{\rho}$ does not change. 
Using this we find 
\begin{equation}
{d \ln A_0 \over d \delta_b}={\int (1+b_L(\nu)) \nu f(\nu) M d\ln \nu \over 
\int  \nu f(\nu) M d\ln \nu}=\langle (1+b_L) \rangle. 
\end{equation}
So the logarithmic slope of $A_0$ with respect to a long wavelength modulation is given by the appropriate 
average of the Eulerian bias $b_E=1+b_L$. 

If instead one looks at the logarithmic growth of the amplitude with respect to amplitude $\sigma_8 \propto \sigma(M)$, 
then $d\nu/d\ln \sigma_8=-2\nu$, and so 
\begin{equation}
{d \ln A_0 \over d \ln \sigma_8}=\delta_c {\int b_L(\nu) \nu f(\nu)M d\ln \nu 
\over \int  \nu f(\nu)M d\ln \nu} =\delta_c \langle b_L \rangle.
\end{equation}
Since $d \ln A_0 /d\ln \sigma_8 = 3.9$ we find $d \ln A_0 / d \delta_b = 3.9/1.68+1 = 3.3$. 
The response to the long wavelength mode has thus a lower logarithmic slope of growth relative to $\sigma_8$ and 
is not much larger than the linear regime value $68/21$. This should be contrasted against the response to the 
amplitude change, which goes from $\sigma_8^2$ in the linear regime to $\sigma_8^{3.9}$ in the nonlinear regime. 
Note that this calculation is valid if the density is divided by the global background density, as appropriate 
for weak lensing observations, which are sensitive to the total density. Whenever the density perturbation is 
defined using local mean density these numbers should be reduced by 2. 

Numerical results are shown in the left panel of figure \ref{fig:ssv}, where we show the nonlinear response to $\delta_b$ from 
simulations of \cite{2014PhRvD..89h3519L}, and the corresponding response to a change in $\sigma_8$ that mimics 
$\delta_b$ at low $k$. 
We can still model a change in $\delta_b$
as a quasi-linear term and $(A_0-A_2k^2+A_4k^4)F(k)$. For the quasi-linear term we adopt 
simply the Zeldovich approximation 
model multiplied with the corresponding linear factor of $68/21\delta_b$, and we fit for 
the other three parameters. The result is shown in figure \ref{fig:ssv} and 
provides a reasonable fit to the simulations. Note that we show results with and without 
$d\ln P / d\ln k$ term, against simulations with and without it \citep{2014PhRvD..89h3519L}. 
We find that for 
$\delta_b=0.02$ $A_0$ has changed by 7.4\%, while the quasi-linear term has changed by 6.4\%, so that 
$d\ln P/d\ln \delta_b=3.2$ at low $k$ and $3.7$ around $k \sim 0.5\ h \rm{Mpc}^{-1}$ where $A_0$ dominates. This is 
in a reasonable agreement with the analytic estimate of 3.3. 
For the $\sigma_8$ scaling a change of 6.4\% in the linear term corresponds to 
13\% change in $A_0$. 
The contrast between the two effects is shown in figure \ref{fig:ssv}. The super-sample variance 
is thus not degenerate with $\sigma_8$, so if one can determine both the quasi-linear term and 
$A_0$ term with sufficient accuracy, one can break the degeneracy between the two effects.

How well can one determine $\sigma_8$ in the presence of super-sample variance? 
If we only have information from $A_0$ then the analysis above suggests that one can determine 
$\sigma_8$ to about $(3.7/3.9)0.4\% \sim 0.38\%$ in $1(h^{-1}{\rm Gpc})^3$ volume, about a factor of 2 worse than 
without the super-sample variance. If we have information both from linear regime and from $A_0$ 
dominated regime then we can break the degeneracy between the super-sample variance and $\sigma_8$. 
The extent to which this can be achieved depends on how well we can measure the amplitude in the 
linear regime: to reach 0.4\% accuracy we would need to measure all the modes up to $k \sim 0.2\ h \rm{Mpc}^{-1}$
in a $1(h^{-1}{\rm Gpc})^3$ volume, which seems possible to achieve. Moreover, we note that a change in 
curvature cannot be modeled well with just a change in linear term and $A_0$, higher order terms also 
change significantly. Even though we argue below that these effects are degenerate with baryonic effects, 
this degeneracy may be broken in this situation given how different these effects are and given that 
there is a lot of information present at high $k$. 
In summary, the amplitude of fluctuations in 
a $1(h^{-1}{\rm Gpc})^3$ volume can be determined to an accuracy of 0.4\% if the super-sample 
variance cannot be determined, which can be reduced by a factor of 2 if the degeneracy between the super-sample 
variance and $\sigma_8$ amplitude can be broken. 

Instead of including the super-sample variance 
effect in the covariance matrix one can include it as an additional curvature parameter that 
one can marginalize over. The parameter is $\delta_b$ and its prior should be a Gaussian with a
zero mean and rms variance $\sigma_V$ determined by the survey window (see \cite{2013PhRvD..87l3504T,2014MNRAS.441.2456T} for predictions for 
simple survey geometries). The response of the power spectrum to the long wavelength 
$\delta_b$ parameter should be 
\begin{equation}
\delta P=\left({47 \over 21}P_{\rm Zel}-{1 \over 3}{dP \over d\ln k}+\left[3.7A_0-3A_2k^2+2.5A_4k^4 \right]F(k)\right)\delta_b,
\end{equation}
where $P_{\rm Zel}$, $A_0$, $A_2$ and $A_4$
are the values of the fiducial model around which we are exploring the super-sample variance effect. 
For  example, in a MCMC chain this would be 
the model one is testing at a given chain position. 
We found that the fit to the simulations must include $A_2$ and $A_4$ terms and that the fit is only 
valid to $k \sim 0.7\ h \rm{Mpc}^{-1}$. Note that the change of $A_2$ and $A_4$ relative to $A_0$ is 
similar to that of amplitude change in equation \ref{slopea024}.

\section{Effects of baryons}
\label{sec:baryons}

\begin{figure*}
\centering
\subfigure{\includegraphics[width=0.48\textwidth]{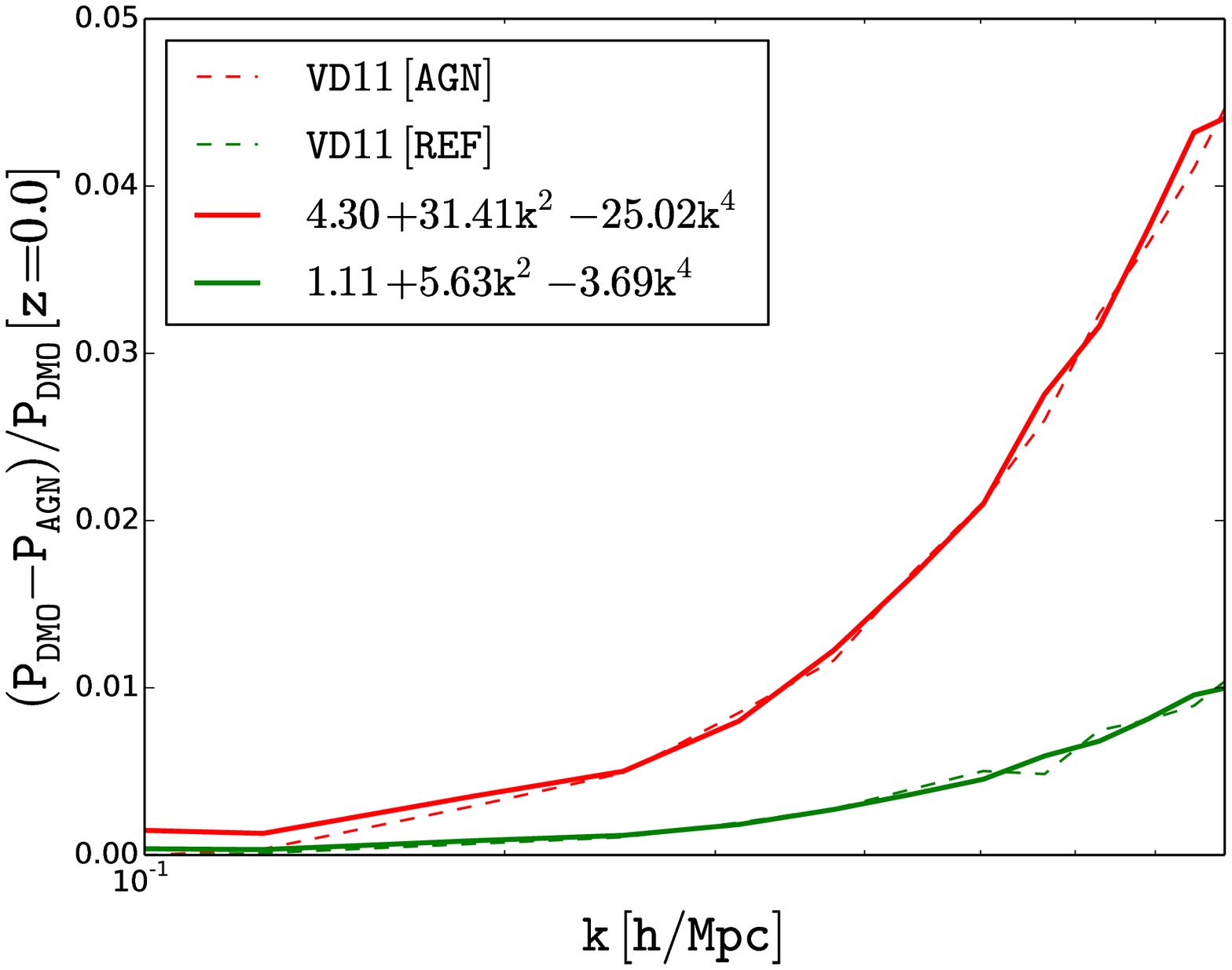}}
\subfigure{\includegraphics[width=0.48\textwidth]{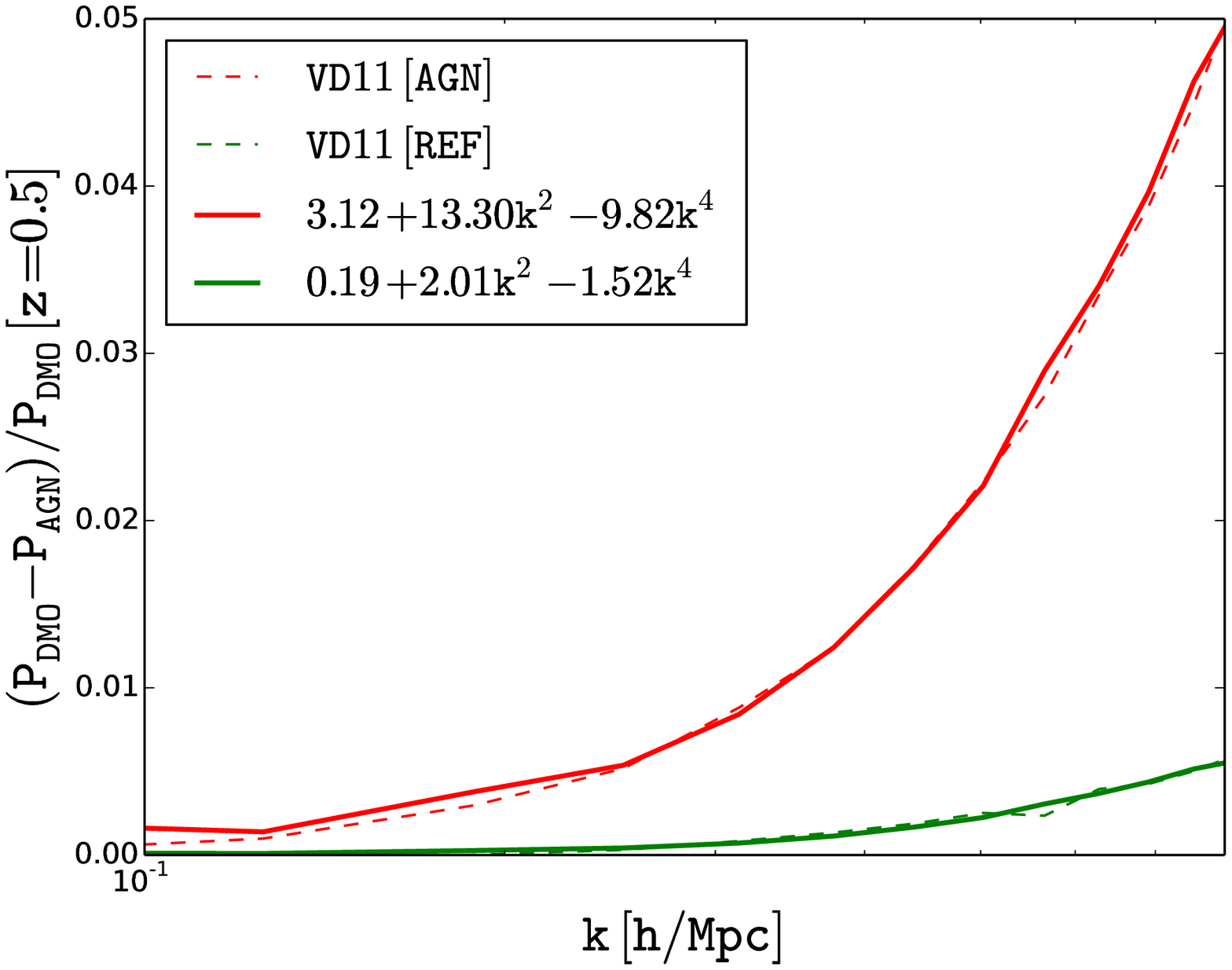}}
\subfigure{\includegraphics[width=0.48\textwidth]{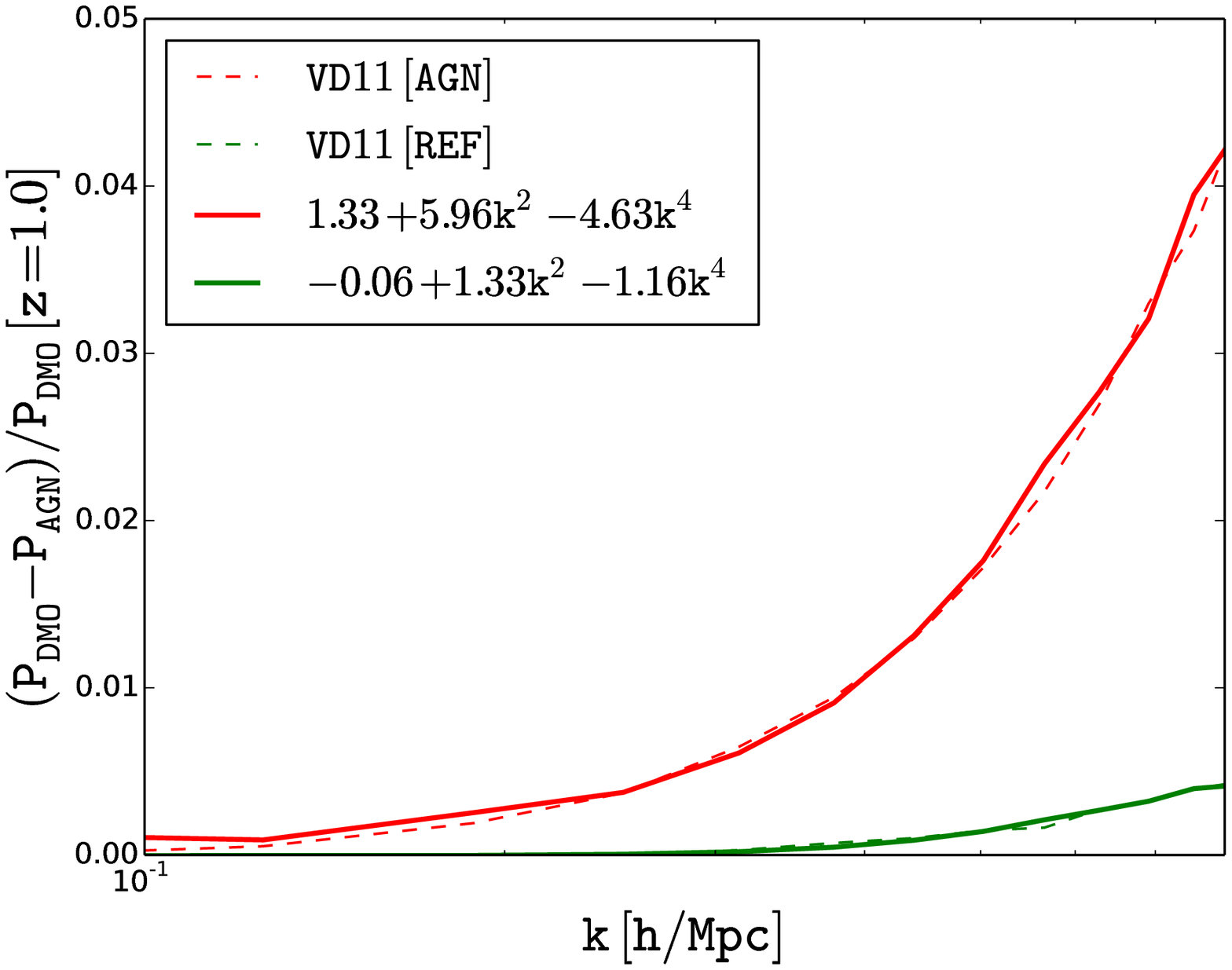}}
\subfigure{\includegraphics[width=0.48\textwidth]{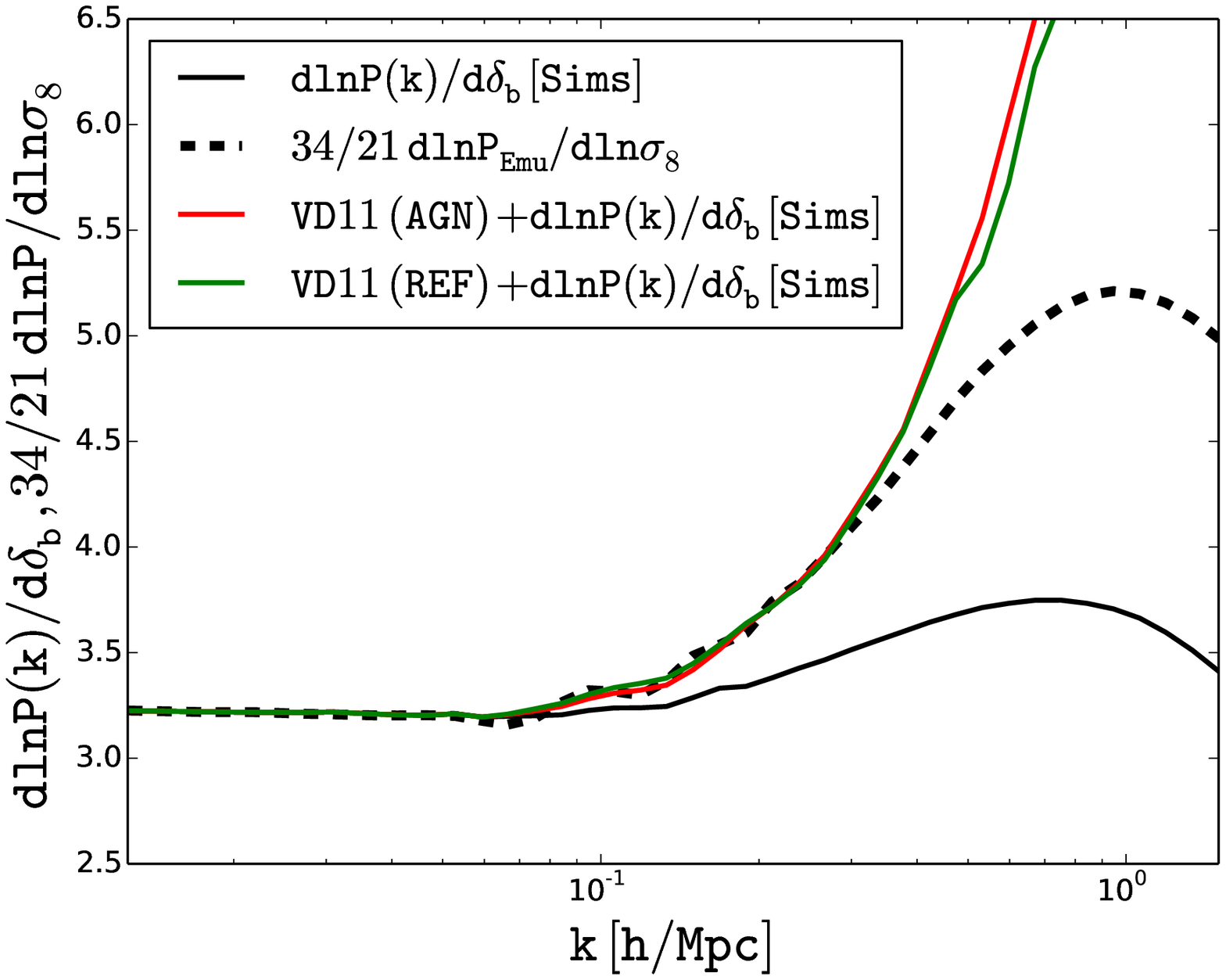}}

\caption{The first three panel (in reading order) are the relative difference between DM only model and AGN (red-dashed line) and REF (green-dashed line) from van Daalen et. al. 2011 (VD11) at redshift 0.0 (top-left), 0.5 (top-right) and 1.0 (bottom-left). REF model contains the baryonic physics without any AGN feedback model. Solid lines (red and green) are the corresponding best fit $\delta A_0-\delta A_2k^2 + \delta A_4k^4$ as explained in section \ref{sec:baryons}. Bottom-right panel shows the derivative of the matter power spectrum with respect to a change in background density (2$\%$) in solid-black from Li et. al. 2014, and with respect to a change in 
amplitude using prediction from emulator (thick dashed-black line).} 
\label{fig:baryons}
\end{figure*}

Baryonic effects inside the dark matter halos change the matter power spectrum relative to the dark matter 
alone and these effects must be incorporated into the analysis, otherwise they can lead to substantial bias 
in the cosmological parameter estimation \citep{2011MNRAS.417.2020S,2013MNRAS.434..148S}. 
Baryonic effects can come in different forms. First is 
simply the fact that gas distribution inside dark matter halos is distributed differently than the dark matter, 
because gas is hot and has significant pressure. As a result, gas has a core at the center of the cluster, leading 
to reduced clustering strength on small scales. Second effect is baryon cooling, which causes gas to cool and 
condense into galaxies at the dark matter halo centers. This leads to an enhancement of the clustering relative 
to pure dark matter case. Baryons can also be pushed out of the halo centers by processes such as 
supernova and AGN feedback, which can in some cases push the gas quite far out. Furthermore, in all of these
examples dark matter may also be redistributed as a consequence of the baryons either condensing onto the 
halo centers or being pushed out. For example, 
for baryonic cooling onto a galactic disk this process is known as adiabatic contraction \citep{1984Natur.311..517B}.  

From the halo model point of view the main effect of the baryons is the redistribution of the gas, and possibly 
dark matter, inside the halos. This can be qualitatively described as the change in the 
scale radius $R_s$. The total mass of the halo $M$ is unchanged, since these baryonic processes do not push the 
gas or the dark matter far out of the virial radius of the halo such that the halo mass would be affected. 
As a consequence, we expect that $A_0$ parameter is essentially unchanged, while $A_2$, $A_4$ etc. will change 
during the baryonic redistribution of matter. 

To investigate this further we used simulation based matter power spectra from \cite{2011MNRAS.415.3649V} to compute
the effects of baryons on the coefficients $A_0, A_2$ and $A_4$. In particular, we use the dark matter only 
 and the supernova and AGN feedback models, corresponding to hydrodynamical simulations with supernova or
AGN feedback model. It was argued that the latter is needed to reproduce 
cluster observations such as X-ray luminosity temperature 
relation \citep{2010MNRAS.406..822M}. We use the AGN model as the main model 
since it provides the largest effects, but we also explore reference supernova feedback model from 
\cite{2011MNRAS.415.3649V}. 
Baryon corrections to the matter power spectrum from AGN feedback model 
exceed one percent level for $k> 0.3 \ h \rm{Mpc}^{-1}$ \citep{2011MNRAS.415.3649V}. 
We use the results at three different redshifts: 0.0, 0.5 and 1.0. In figure \ref{fig:baryons} we try to fit the difference between the pure dark matter and AGN model, $\rm{P_{DMO} - P_{AGN}}$, or reference supernova feedback
model $\rm{P_{DMO} - P_{REF}}$, with the model
$\delta A_0 + \delta A_2k^2 + \delta A_4 k^4$, 
to estimate the changes in these coefficients due to baryons at
each redshift (note that since the changes are only important at high $k$ we can set 
$F(k)=1$). We fit these models over the $k$ range between 0.2 and 0.8 $\ h \rm{Mpc}^{-1}$. Figure \ref{fig:baryons} 
shows the best fit models, which are a good fit to the simulations over this range. 
We also calculated these coefficients for the cosmology assumed in this paper using the results from figure \ref{fig:A_sigma}.

We find that for the AGN model the relative change in $A_0$ is about $0.5-1\%$, depending on the redshift, 
whereas the changes in $A_2$ and $A_4$ are about $4-7\%$ and 4-8 $\%$, respectively. 
If we assume no change in $A_0$ the fit is a bit worse and
the change in $A_2$ and $A_4$ is larger. 
This confirms that the coefficient $A_0$ is quite indifferent to baryonic effects, while
$A_2$ and $A_4$ are significantly more contaminated. 
The change is positive. This is expected since AGN feedback expands 
the gas and makes the scale radius $R_s$ larger. It is less obvious why $A_0$ should increase when gas is being pushed 
outwards, but the effect on $A_0$ is small and it could also be driven by the numerical fitting procedure. 
If we assume that the baryonic uncertainty is at the level suggested by these AGN models, then using 
equation \ref{eqn:covariance} the 
corresponding uncertainty on $\sigma_8$ will be $0.5-1\%/3.9 \sim 0.1-0.2\%$ from $A_0$, and about an 
order of magnitude larger from $A_2$ and $A_4$. Given that the difference between AGN and DM models is probably 
an overestimate of the error associated with the baryonic effects our analysis
suggests that these effects can be effectively 
marginalized over without any loss of cosmological information from $A_0$. 
We also note that other baryonic feedback models from \cite{2011MNRAS.415.3649V}, such as the 
reference model, while giving a lower amplitude of 
the effect, have very similar $k$-dependence, as can be seen from figure \ref{fig:baryons}.

Above we argued that super-sample variance effect should not be treated as a variance but as a separate 
parameter that can be determined from the data. Using linear theory and $A_0$ may not contain enough information 
to break the degeneracy between the amplitude $\sigma_8$ and super-sample variance. Using higher $k$ information may be 
more promising, since the two effects also have very distinctive signatures on $A_2$, $A_4$ etc. Since our 
model expansion to $A_4$ only works to $k \sim 0.7\ h \rm{Mpc}^{-1}$ we explore this question numerically. 
In figure \ref{fig:baryons} bottom-right we plot the 
super-sample effect and amplitude effect such that they are degenerate at low $k$, while also adding the
baryonic effect 
such that it is degenerate with change of amplitude up to $k \sim 0.3\ h \rm{Mpc}^{-1}$. 
We see from the figure \ref{fig:baryons} that above $k \sim 0.3\ h \rm{Mpc}^{-1}$ 
the degeneracy is broken: the effects of $\sigma_8$ and $\delta_b$ are smaller compared to 
the effect of AGN 
feedback, which
continuous to increase with $k$, because gas is being pushed out on small scales, suppressing all small 
scale clustering. While this analysis is only restricted to a specific form of baryonic effects and is less robust 
than the other analyses in this paper, it suggests that one may be able to break the degeneracy between the 
baryonic effects, cosmological parameters such as amplitude, and super-sample variance, using high $k$ information.

\section{Discussion and conclusions}\label{sec:discussion}

In this work we propose a model of the matter power spectrum using the Zeldovich approximation
power spectrum as the 2-halo term and even powers of $k$ expansion of the 1-halo term, compensated 
on large scales to satisfy mass conservation, with coefficients 
calibrated on simulations. The leading order 1-halo term is $k^0$ term amplitude $A_0$, which
in the halo model can be determined as a mass dependent integral over the halo mass function.
Simulations predict $A_0 \propto \sigma_8^{3.9}$, and the halo model is only able to reproduce this at low 
redshifts. 
The amplitude of $A_0$ is related to the cluster 
abundance method, where one counts clusters above a given mass, which also depends on the halo mass
function and has a similarly 
steep dependence on $\sigma_8$. It is also related to SZ power spectrum 
scaling, which is dominated by the 1-halo term and scales
as $\sigma_8^7$ \citep{2002MNRAS.336.1256K}, because the SZ signal from individual clusters 
scales as $M^{5/3}$ rather than 
halo mass $M$ and it is a projection over line of sight, leading to a steeper dependence on $\sigma_8$.
Our analysis thus explicitly connects the cluster abundance method to the amplitude of the leading nonlinear 
correction to the matter power spectrum, and shows the two use similar information. As a consequence, these two 
methods cannot be combined independently if the dominant errors are Poisson or large scale structure fluctuations. 

Using the first 3 coefficients of expansion 
we accurately predict variations of basic cosmological parameters up to $k\sim 0.7\ h \rm{Mpc}^{-1}$, including amplitude 
$\sigma_8$, matter density $\Omega_m$, Hubble parameter $H_0$, primordial slope $n_s$, 
equation of state $w_0$ and even neutrino mass $\sum m_{\nu}$. In all cases our model predicts well the BAO 
smoothing, a consequence of using the 
Zeldovich approximation rather than linear theory for the 2-halo term. 

We present a very simple 
model for the covariance matrix of matter power spectrum (equation \ref{cov}). 
We stress that the covariance matrix depends on the simulated volume both in linear and 
nonlinear regimes, so a direct comparison between covariance matrices from 
different simulations needs to account for this. 
In this model the large scale variance is dominated by the 
sampling variance, while 
on small scales where $A_0$ dominates the dominant term is the Poisson sampling of the halos. 
Using the halo mass function  of \cite{2008ApJ...688..709T} to predict the latter gives about 20-30\% higher value than 
fitting with simulations of \cite{2014PhRvD..89h3519L}, which we consider a good agreement given the 
inaccurate nature of halo mass function fits in the high mass regime.
Using this value we 
show that our model gives a remarkable agreement with 
the simulations of \cite{2014arXiv1406.2713B}, where 12288 simulations of $656 h^{-1}{\rm Mpc}$ box size were run to 
construct a covariance matrix. 
We use our Poisson model to compute the convergence rate of the covariance matrix and find that simulated 
volumes of 500-5000$(h^{-1}{\rm Gpc})^3$ are needed to converge at 1\% level. This explains why our model without any free parameters 
does not reproduce covariance matrix 
\cite{2012MNRAS.423.2288H}, because the total volume used in \cite{2012MNRAS.423.2288H} 
was only 1.6$(h^{-1}{\rm Gpc})^3$, and has thus not converged with high enough accuracy. 
Changing the parameter $\sigma_{A_0}/A_0$ from 
the predicted 0.09 to 0.15 we obtain a perfect agreement. 

Using this model we argue that most of the cosmological information about the amplitude 
is in $A_0$, which can determine the amplitude $\sigma_8$ to 0.2\% within $1(h^{-1}{\rm Gpc})^3$ volume. 
The higher order coefficients $A_2$, $A_4$ etc. 
are less sensitive to $\sigma_8$ and have a larger variance.
We discuss the super-sample variance and argue that due to its origin as a curvature effect it
differs from the amplitude rescaling and so it 
should be treated as a separate cosmological parameter with a 
prior given by the rms variance on the scale of the survey volume. 
If its degeneracy with the amplitude is not broken then it approximately doubles the errors, so that 
$\sigma_8$ can be determined to 0.4\% within $1(h^{-1}{\rm Gpc})^3$  volume. Note that both of these errors are a lot 
smaller than the currently available constraints, which at best are at 
4\% \citep{2013MNRAS.430.2200K}: observational and modeling errors 
dominate the error budget 
at the moment, but future data sets may be able to reach the levels where super-sampling variance 
or Poisson error will dominate \citep{2012PhRvD..86h3504Y}. 

We also investigate the baryonic effects on the matter power spectrum. We argue that these 
should not change $A_0$ much because of the mass conservation. Indeed, comparison of our model to 
simulations of baryonic effects in \cite{2011MNRAS.415.3649V} suggests that $A_0$ is almost unchanged, while higher order 
coefficients change significantly, because baryonic effects redistribute gas and dark matter 
inside the halos without changing the overall halo mass. 
We advocate that marginalizing over higher order expansion coefficients 
should immunize against baryonic effects without much loss of information. We explore the degeneracy 
between the amplitude, super-sample variance and baryonic effects, finding that it can be broken using 
information above $k \sim 0.3\ h \rm{Mpc}^{-1}$. 

Our results suggest that analytic modeling of dark matter clustering provides important insights 
even in the era of large simulations. 
It offers a promising 
venue not only for an accurate power spectrum description, but also for the covariance matrix modeling, 
for optimal extraction of information from the data, and for description of baryonic effects. 
We have shown that in the context of covariance matrix calculations our model is likely to be more reliable 
than simulations with insufficient total volume. 
However, more work remains to be done before it can be applied to the weak lensing observations. 
For example, 
in this paper we focused on the dark matter clustering description in terms of its
power spectrum. If one wants to apply the method
to the weak lensing observations one needs to perform the line of sight projections of the model onto the weak
lensing power spectrum $C^{\kappa \kappa}_{l}$, where $\kappa$ is the convergence which can be 
written as a projection of the density along the line of sight. Projecting powers of $k$ simply gives the same 
powers of $l$, so if the projection kernels are narrow, as would be the case for weak lensing 
tomography, the analysis remains essentially 
unchanged, except for the fact that weak lensing probes matter density rather than density 
perturbations, so convergence is also multiplied by an overall mean 
matter density. If the projection kernels are broad and there are significant 
contributions from nearby structures for which $k>0.7\ h \rm{Mpc}^{-1}$ projects to a low $l$, then one
needs to assess these effects and improve the model to account better for the high $k$ contributions. 
Similarly one also needs to project baryonic effects and covariance matrix. 
This program is feasible and if implemented it will give a completely analytical 
description of the weak lensing power spectrum and its covariance matrix
without any need to use simulations. 

\section{Acknowledgements}
We thank Z. Vlah for extensive discussions and for providing the Zeldovich power spectrum code, F. Schmidt and M. Takada for
useful comments, Y. Li for electronic form of the plots of \cite{2014PhRvD..89h3519L}, M. van Daalen for electronic form of the plots of \cite{2011MNRAS.415.3649V}, L. Blot for providing us data from their simulations, and J. Harnois-Deraps and U. Pen for interpretation of their covariance matrix.
U.S. is supported in part by the NASA ATP grant NNX12AG71G. I.M. would like to thank the hospitality of LBNL.

\bibliographystyle{mn2e}
\def\apj{ApJ}
\def\apjl{ApJL}
\def\aj{AJ}
\def\mnras{MNRAS}
\def\aap{A\&A}
\def\nat{Nature}
\def\pasj{PASJ}
\def\prd{PRD}
\def\physrep{Physics Reports}
\def\jcap{JCAP}
\def\araa{ARAA}
\bibliography{ms.bib}
\newpage

\end{document}